\documentclass[article, shortnames, nojss]{jss}

\usepackage{thumbpdf, lmodern}

\usepackage{amsmath, amssymb, caption, subcaption, multirow}

\usepackage[symbol]{footmisc}
\usepackage{perpage}
\MakePerPage{footnote}

\newcommand{\ud}{\mathrm{d}}

\author{Andrew Iskauskas \\ Durham University \And Ian Vernon \\ Durham University \And Michael Goldstein \\ Durham University \AND Danny Scarponi \\ London School of Hygiene \\ and Tropical Medicine \And Trevelyan J. McKinley \\ University of Exeter \And Richard G. White \\ London School of Hygiene \\ and Tropical Medicine \AND Nicky McCreesh \\ London School of Hygiene \\ and Tropical Medicine}
\Plainauthor{Andrew Iskauskas et al.}

\title{Emulation and History Matching using the \pkg{hmer} Package}
\Plaintitle{Emulation and History Matching using the hmer Package}
\Shorttitle{The hmer Package}

\Abstract{
Modelling complex real-world situations such as infectious diseases, geological phenomena, and biological processes can present a dilemma: the computer model (referred to as a simulator) needs to be complex enough to capture the dynamics of the system, but each increase in complexity increases the evaluation time of such a simulation, making it difficult to obtain an informative description of parameter choices that would be consistent with observed reality. While methods for identifying acceptable matches to real-world observations exist, for example optimisation or Markov chain Monte Carlo methods, they may result in non-robust inferences or may be infeasible for computationally intensive simulators. The techniques of emulation and history matching can make such determinations feasible, efficiently identifying regions of parameter space that produce acceptable matches to data while also providing valuable information about the simulator's structure, but the mathematical considerations required to perform emulation can present a barrier for makers and users of such simulators compared to other methods. The \textbf{hmer} package provides an accessible framework for using history matching and emulation on simulator data, leveraging the computational efficiency of the approach while enabling users to easily match to, visualise, and robustly predict from their complex simulators.
}

\Plainkeywords{emulation, history matching, calibration, R}

\Address{
  Andrew Iskauskas\\
  Department of Mathematical Sciences\\
  Mathematical Sciences \& Computing Science Building\\
  Durham University\\
  Upper Mountjoy Campus\\
  Stockton Road\\
  Durham DH1 3LE \\
  E-mail: \email{andrew.iskauskas@durham.ac.uk}\\

}

\graphicspath{{images/}}

\begin{document}

\section{Introduction: emulation and history matching}

In many scientific disciplines, complex computer models, or \emph{simulators}, are necessary to comprehend or explore the behaviour of physical systems. Examples range from biological simulations at the microscopic level \citep{vernon2018bayesian} to physical simulations of galaxy formation in the observable universe \citep{bower2010galaxy}; of particular interest and relevance is the application of computer simulators to models of infectious disease and epidemics. Despite their disparate subjects most, if not all, of these models share similar characteristics: by virtue of the complexity of the system being represented the simulators themselves are often necessarily complex, requiring a large parameter space and a lot of computational time to run. If one wishes to use the simulator to match to observed data (for instance to gain insight into the initial conditions of the formation of a particular galaxy, or the characteristics of a particular strain of an infection), this can be an arduous task where the region of parameter space with non-negligible support under the data is small, hard-to-find, and often requires an infeasible number of simulator evaluations. Even if a match to observed data can be found, we may not be able to determine with any certainty whether the found parameter set is truly representative of the state of the real-world system due to the various uncertainties present.

We deal with these drawbacks with two methodologies. The first, emulation, provides a means by which we can statistically represent the output of the simulator given a sample of runs from it. The emulators that we construct are fast to evaluate at unseen parts of parameter space, and their statistical nature allows us to explicitly and rigorously encode the uncertainty around predictions at any point in parameter space. This alone can be useful for understanding the structure of the simulator as well as offering a gateway to a deep analysis of the uncertainty in a (necessarily imperfect) representation of a physical system given by the simulator. In order to address the issue of finding acceptable matches to observed data, we couple the emulation to a process of `history matching': a means of iteratively removing unacceptable parts of parameter space so that, by complementarity, we find the full space of acceptable parameter combinations. The conjunction of these two techniques allows us to explore the parameter space systematically and quickly, minimising the number of computationally expensive simulator evaluations required to identify all possible combinations of input parameters that could result in a match to observational data.

The history matching and emulation approach has been employed in a variety of epidemiological systems, including for models of HIV \citep{andrianakis2017efficient, andrianakis2017history} and the implementation of anti-retroviral therapy therein \citep{mccreesh2017improving, mccreesh2017universal}, models of tuberculosis and HIV across multiple countries across the world \citep{clark2022tbhiv}, and large-scale agent-based models of Covid in the UK \citep{krauss2022june, vernon2022bayesianjune}. Emulation has also been applied successfully in a range of disciplines beyond epidemiology: for example, in astrophysics \citep{higdon2004combining, kaufman2011efficient, vernon2014galaxy}; climate science \citep{castelletti2012general, williamson2013history, edwards2021projected}; engineering \citep{du2021optimization}; and vulcanology \citep{gu2016parallel, marshall2019exploring}. However, the statistical machinery required to efficiently and accurately apply emulation to such simulators presents a barrier to most modelling communities. Techniques such as optimisation, Markov-chain Monte Carlo (MCMC), or Approximate Bayesian Computation (ABC) can be preferred due to their ease of implementation; however, these methods typically require large numbers of simulator evaluations and hence may not be viable, or even computationally tractable, for high-dimensional spaces. Furthermore, such techniques may not find the full space of acceptable matches, making robust inference difficult or even impossible. The creation of the \pkg{hmer} package is designed to remove the conceptual barrier that precludes modellers from using emulation and history matching, providing accessible functionality to perform the history matching procedure while giving the user the flexibility to interact with the mathematical structure as much as they see fit.

There are a number of options in \proglang{R} that perform emulation on complex models, the majority of which focus on Gaussian Process (GP) emulation. Notable examples are \pkg{emulator} \citep{hankin2005emulator} and its successor \pkg{multivator} \citep{hankin2012multivator}, \pkg{stilt} \citep{olson2018stilt} and \pkg{RobustGaSP} \citep{gu2022robustgasp}. While GP emulation is a powerful tool, it can be challenging for a user to adequately specify the full Bayesian prior structure for such an emulator and it includes distributional assumptions that can be hard to justify. In addition, these packages focus (understandably) on the emulation of simulator outputs and not on the tools that would aid a user in understanding and visualising the structure of their simulator via emulation. They also do not leverage the power of history matching, a key tool when the principal expected usage of emulation is to find acceptable fits to data arising from complex simulators of complex real-life processes. In contrast, due to the more flexible framework and simpler prior specifications used, \pkg{hmer} may be a more accessible tool for emulation and, coupled with the power of history matching, provides a more straightforward means of approaching complex calibration problems using emulation.

The structure of this paper is as follows. In Section \ref{sec:hme} we briefly outline the fundamentals of emulation and history matching. In Section \ref{sec:hmer} we detail the functionality of the main components of the \pkg{hmer} package, along with their arguments. In Section \ref{sec:ex} we apply the core functionality, along with some of the visualisation tools, to a toy model to demonstrate its usage. In Section \ref{sec:advanced} we discuss one of the more advanced techniques available to an expert user, particularly variance emulation for stochastic simulators. Finally, in Section \ref{sec:conclude} we conclude and discuss considerations that should be made during the process of emulation.

\section{Bayes linear emulation and history matching}\label{sec:hme}

In this section we briefly outline the mathematical framework of Bayes linear emulation, as well as the algorithmic description of history matching. Multiple works exist that provide a fuller description of emulation; see, for instance, \cite{craig1997pressure, vernon2018bayesian, santner2003design, bowman2016emulation}.

\subsection{Emulation}

Suppose we have a simulation of a real-world process which takes a set of input parameters, described as a vector $x$ of length $d$, and returns a set of $m$ outputs $\{f_i(x)\}_{i=1,\dots,m}$. A (univariate) Bayes linear emulator is a fast statistical approximation of the simulator output, built using a comparatively small set of simulator runs, which provides an expected value for the simulator output at any point $x$ along with a corresponding uncertainty estimate reflecting our beliefs about the uncertainty in the approximation.

Concretely, we may create a prior representation of a simulator output $f_i(x)$ in emulator form as

\begin{equation}\label{eq:emulator}
f_i(x) = \sum_{j=1}^{p_i} \beta_{ij} g_{ij}(x_{A_i}) + u_i(x_{A_i}) + w_i(x).
\end{equation}

Here, $x_{A_i}$, $A_i \subseteq \{1,\dots,d\}$, are the set of `active variables' for output $f_i(x)$; that is, the components of $x$ that are most influential in determining the behaviour of $f_i(x)$. The $g_{ij}$ are $p_i$ known simple functions of $x_{A_i}$, with the $\beta_{ij}$ the corresponding coefficients; together these two terms define a regression surface encoding the global behaviour of the output. $u_i(x_{A_i})$ is a second-order weakly stationary process which captures residual variation in $x_{A_i}$ and can be seen as governing the local behaviour of the simulator output. We make the assumption that $u_i(x_{A_i})$ is zero-mean and \textit{a priori} uncorrelated with the regression coefficients $\beta_{ij}$. We further assume the following covariance structure for $u_i(x_{A_i})$:

$$\COV[u_i(x_{A_i}), u_i(x^{\prime}_{A_i})] = (1-\delta_i) \sigma^2_i c(x_{A_i}, x^{\prime}_{A_i}),$$

where $c(x,x^{\prime})$ is a suitable correlation function \citep[common examples can be found in][Ch.~4]{rasmussen2003gaussian} and $\delta_i \in [0,1]$. The `nugget' term, $w_i(x)$, represents the effects of the remaining `inactive' inputs; we again assume this is zero-mean and uncorrelated to $\beta_{ij}$ and $u_i(x_{A_i})$, and that

\begin{equation}\label{eq:corr}
\COV[w_i(x), w_i(x^{\prime})] = \delta_i \sigma^2_i I_{x=x^{\prime}},
\end{equation}

where $I_{x=x^{\prime}}$ is an indicator function with $I_{x=x^{\prime}}=1$ if $x=x^{\prime}$ and $0$ otherwise.

Before we can update the emulated structure with respect to data, we need to complete the \textit{a priori} specification for the random quantities $\beta_{ij}$, $u_i(x_{A_i})$, and $w_i(x)$. This can be done in a variety of ways: for instance, if one is willing and able to specify full distributions for these quantities, we could then use maximum likelihood or maximum \textit{a posteriori} (MAP) estimates to determine plug-in estimates for their hyperparameters \citep{andrianakis2012effect}, or use cross-validation \citep{maatouk2015cross}. We may not be able (or willing) to make full distributional specifications for these quantities, whether due to a lack of prior knowledge required for such a specification or a lack of faith in any such specification. It is rare, however, that we have similar reservations about the second-order specification of such a system (that is, expectations and covariances), and the Bayes linear framework requires only these quantities. We therefore leverage this framework and require, with the assumptions listed already, specification of the expectation and covariance of the regression coefficients $\E[\beta]$ and $\VAR[\beta]$, as well as the quantities that furnish the covariance structure of $u_i(x)$ and $w_i(x)$; namely $\delta_i$, $\sigma_i$, and the hyperparameters of the correlation function $c(x, x^{\prime})$. These can be determined using a full \textit{a priori} specification or by using pragmatic plug-in estimates \citep{santner2003design, kennedy2001bayesian, rasmussen2003gaussian}. With these quantities defined, we are in a position to update our knowledge in light of data using the Bayes linear framework, which we now describe.

Let us imagine that we have a collection of runs obtained from running the simulator at a series of points $(x^{(1)}, x^{(2)}, \dots, x^{(n)})$, resulting in a collection of simulator outputs 
$$D_i=\left(f_i(x^{(1)}),\dots,f_i(x^{(n)})\right).$$ 
The Bayes linear emulator output for $f_i(x)$ at a new point $x$ is given by the Bayes linear update formulae \citep{goldstein2007bayes}:

\begin{align}\label{eq:blupdate}
\E_{D_i}[f_i(x)] &= \E[f_i(x)]+\COV[f_i(x),D_i]\VAR[D_i]^{-1}(D_i-\E[D_i]), \\
\VAR_{D_i}[f_i(x)] &= \VAR[f_i(x)]-\COV[f_i(x),D_i]\VAR[D_i]^{-1}\COV[D_i,f_i(x)],
\end{align}

as well as the covariance between the outputs at two points $x,\,x^{\prime}$

$$\COV_{D_i}[f_i(x),f_i(x^{\prime})]=\COV[f_i(x),f_i(x^{\prime})]-\COV[f_i(x),D_i]\VAR[D_i]^{-1}\COV[D_i,f_i(x^{\prime})].$$

\begin{figure}[!h]
\centering
\includegraphics[width=0.8\textwidth]{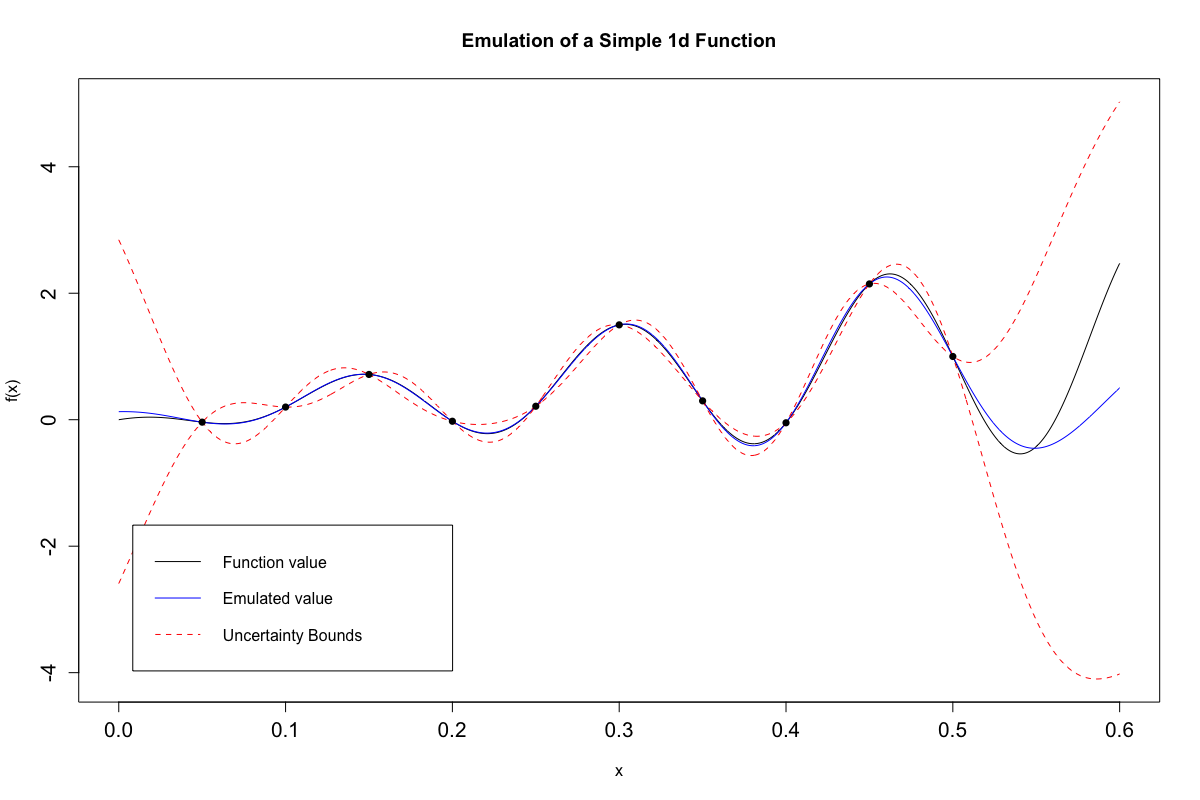}
\caption{An example of emulation of a simple one~dimensional example.}
\label{fig:oned}
\end{figure}

The emulator expectation given by \eqref{eq:blupdate} provides a prediction for $f(x)$ at an unevaluated point $x$, while the emulator variance in (4) provides us with the corresponding uncertainty of this prediction. Both of these quantities are extremely fast to evaluate, requiring little more than matrix multiplication, so the emulators allow an extensive exploration of the function's behaviour over the input space. A one-dimensional example is shown in Figure~\ref{fig:oned}, where the true function can be evaluated: the black line represents the true value, while the blue line is the corresponding emulator prediction. Red dotted lines represent $2\sigma$ bounds as determined by the emulator uncertainty, and training points $D$ are denoted by black dots. This example is included within the \pkg{hmer} vignettes:

\begin{CodeInput}
R> vignette("low-dimensional-examples", package = "hmer")
\end{CodeInput} 

We note a few general features of emulation with reference to this example. At training points, the emulator exactly reproduces the simulator output with zero variance\footnote{This is true of any deterministic emulation -- this is not true of stochastic emulation, as we will see in Section~\ref{sec:advanced}.}. Away from the training points, the emulator's interpolation and extrapolation follows the simulator values well, with appropriate uncertainty as we move further from known points. The true value does not extend beyond the bounds of the emulator uncertainty, even in the regime where extrapolation is necessary.

We may briefly note that, were we to assume normal and Gaussian process priors for $\beta$ and $u(x)$, respectively, then the approach here is almost identical to Gaussian process emulation \citep{conti2009gaussian}. However, Gaussian process emulation requires invoking additional distributional assumptions that may be difficult to justify, force stricter and more complicated diagnostics to be satisfied, and fundamentally affect the final inference. This demonstrates the advantage of Bayes linear emulation: in requiring only second-order specifications, we are somewhat removed from the associated difficulty and pitfalls inherent in a distributional approach. Finally, the simplicity of the Bayes linear framework allows tractable consideration of more complex applications of emulation, including multi-level or hierarchical emulation \citep{cumming2009small}.

Despite its simple structure, it can still be difficult for a model expert to make the second-order determinations with reasonable certainty. In Section~\ref{sec:emfromdata} we discuss how the \pkg{hmer} package creates the prior specifications and determines hyperparameters (for example those in the correlation function $c(x,x^{\prime})$), in order to create robust emulators.

\subsection{The implausibility measure and history matching}\label{sec:hm}

One of the main advantages of the emulation framework described above is speed compared to the simulator that it approximates. An emulator evaluation at a previously unseen point is often orders of magnitude faster than the complex simulator from which it is derived -- in applications we have used, the speed increase of emulation compared to simulation is between $10^2$ and $10^{10}$ time faster. We can therefore use an emulator to intuit the structure and behaviour of the simulated output over the parameter space, without resorting to large numbers of computationally expensive simulator runs. The statistical nature of the emulators means that any such intuition is made in the presence of the uncertainty of the emulator at a given point, and this must be taken into account. However, a common aim when using complex simulators is to answer the following question:

\emph{Given observed data corresponding to a simulator output, what combinations of input parameters could give rise to output consistent with this observation?}

Of course, there are a variety of approaches to matching complex simulators to data, from optimisation to Monte Carlo methods such as Data-augmented Markov chain (DA-MCMC) or Sequential (SMC) methods \citep{gibson1998estimating, o1999bayesian, jewell2009bayesian}, to Approximate Bayesian Computation (ABC) \citep{mckinley2018approximate, mckinley2009inference, toni2009approximate}, to full Bayesian inference \citep{kennedy2001bayesian}. The benefits and drawbacks of each are varied; primary in our consideration here is that in any technique we apply we should best leverage the structure of our emulators to find the full space of acceptable matches. To that end, we use the history matching approach \citep{craig1997pressure}, which aims to find the space of acceptable parameter sets via complementarity. For further discussion and comparison with other approaches, see \cite{bower2010galaxy, mckinley2018approximate, vernon2018bayesian}. We first must consider the link between an observation of a real-world process to an emulator before we describe the history matching procedure.

Let us denote the real physical process which one output of the complex simulator represents by $y_i$. Observations of this physical process are almost certainly made imperfectly (for example, in epidemiological models it is common to suspect that case numbers for a disease are subject to under-reporting issues). We link the observation $z_i$ to the physical process via

$$z_i = y_i + e_{i},$$

where $e_{i}$ is a random quantity reflecting the uncertainty about the accuracy of our observation. Similarly, we should not expect that the simulator is a perfect representation of the real-life process it represents; the simulator output $f_i(x)$ can be linked to the physical process via

$$y_i = f(x_i^\star) + \epsilon_{i},$$

where $\epsilon_{i}$, the structural model discrepancy, is a random quantity representing our uncertainty about the imperfections of the simulator \citep{goldstein2013assessing, brynjarsdottir2014learning, bower2010galaxy}. Here we follow the ``best input'' approach, which suggests that there exists a value $x^{\star}$ which best represents the real physical system \citep{goldstein2006bayes}. We already have a well-defined means by which to link the emulated output to the simulator, at least to second-order, due to the inherent uncertainty quantification that the emulator provides. This set of uncertainties provides a chain allowing us to link the emulator prediction directly to the observation. We frequently assume that $e_i$ and $\epsilon_{i}$ are zero-mean, constant variance quantities that are uncorrelated with each other and with the emulated output; the generalisation is straightforward in the equations that follow.

This uncertainty structure allows us to approach the problem of matching in a markedly different way to, e.g., optimisation. Rather than seeking points whose simulator outputs are likely to be good fits to the observational data, we instead focus on removing parts of parameter space that are highly unlikely to give rise to good fits, even accounting for the combined uncertainties linking the observation to the emulator. By systematically removing bad parts of parameter space and improving the emulator predictions over the remaining `non-bad' parameter space, we can arrive by complementarity at the complete space of acceptable parameter combinations relative to the observational data.

Concretely, we define an \emph{implausibility measure} \citep{vernon2014galaxy} for observation $z_i$

$$I_i^2(x) = \frac{(\E_{D_i}[f_i(x)]-z_i)^2}{\VAR_{D_i}[f_i(x)]+\VAR[e_i]+\VAR[\epsilon_{i}]}.$$

If $I(x)$ is `large'\footnote{The question of what we mean by `large' or `small' is a good one. We may appeal to Pukelsheim's $3\sigma$ rule \citep{pukelsheim1994three}, which suggests that for a unimodal continuous distribution $95\%$ of the probability mass is within $3\sigma$ of the mean, to suggest a good starting point for a cut-off is $I=3$.} then we may think it unlikely that we would obtain an acceptable match to observed data were we to run the simulator at the point $x$; in this case, we term the point $x$ \emph{implausible}. Conversely, if $I(x)$ is `small' then we cannot rule out the possibility that $x$ would give rise to a good match to data; $x$ is deemed \emph{non-implausible}, or \emph{not-yet-ruled-out (NROY)}. Note that a point can be deemed non-implausible either because the emulator expectation $\E[f_i(x)]$ is close to the observation $z_i$, suggesting a good fit, or because the uncertainties (particularly $\VAR[f_i(x)]$) are large, suggesting that investigation of that part of parameter space using additional simulator runs would allow us greater insight into the structure of the non-implausible space. A point is deemed implausible only if, even accounting for the uncertainties in the simulator, observation, and emulator, a simulator evaluation at this point is very unlikely to provide a match to the observational data.

For multiple outputs we may also define a combined implausibility measure; a natural approach is to require all implausibility measures across all outputs to be satisfied, resulting in a maximum implausibility measure for $m$ outputs:

$$I_M(x) = \max_i\{I_i(x)\}_{i=1,\dots,m}.$$

Other obvious, and less restrictive, measures follow immediately: the second-maximum implausibility is defined as

$$I_{2M}(x) = \max_i\{I_i(x)\setminus I_M(x)\}$$

and so on. Other options are available - for example, the definition of a multivariate implausibility measure \citep{bower2010galaxy} which can capture some of the correlations between outputs; these will be included in a subsequent update of the \pkg{hmer} package.

\begin{figure}[!h]
\centering
\includegraphics[width=0.6\textwidth]{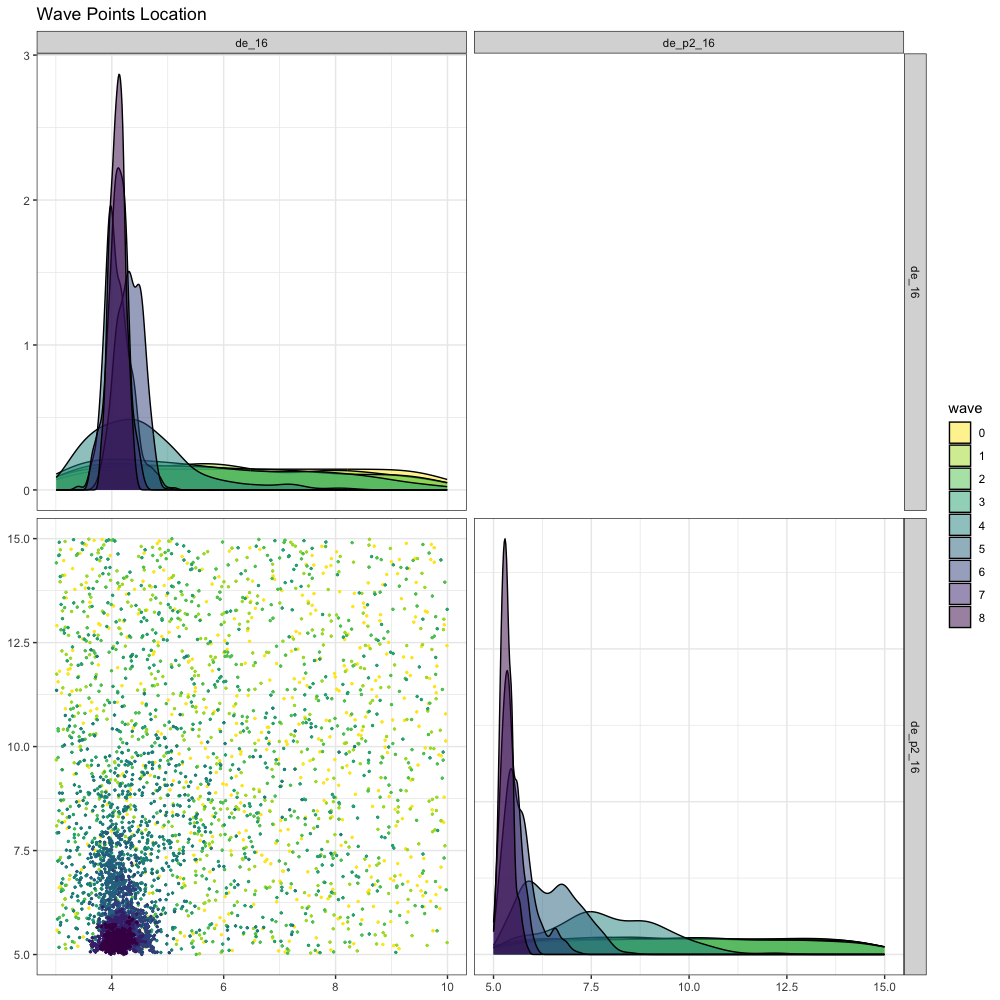}
\caption{A schematic demonstration of the history matching process, over $8$ waves of emulation. The plot is generated using \code{wave\_points()} in the \pkg{hmer} package.}
\label{fig:examplewaves}
\end{figure}

The history matching algorithm proceeds as follows. We apply a series of iterations, called waves, which discard regions of the parameter space at each wave. At wave $k$, a set of emulators are constructed for a collection of outputs, $Q_k$, based on a representative sample of points and their corresponding simulator evaluations from wave $k-1$ \footnote{Note that $Q_k$ need not be the entire set of outputs of interest --- especially at early waves, we may select an informative (and straightforward to emulate) subset of them with a view to including others at later waves. The complementarity of the history matching procedure ensures that this is a valid approach, and allows us to initially focus on `stable' outputs governed by large-scale behaviour across the parameter space \citep{vernon2018bayesian}.}. These emulators are used to assess implausibility over the space, $\mathcal{X}_k$, that remained at wave $k-1$, discarding those regions deemed implausible to produce representatives of a smaller parameter space, $\mathcal{X}_{k+1}$. These points in turn inform the emulators at wave $k+1$ and so on. The algorithm is laid out below, and a schematic example of the process is shown in Figure~\ref{fig:examplewaves}. The example shown here is drawn from an application of the software to a real-world problem, the details of which can be found in \cite{iskauskas2024HPVsim}.

\begin{enumerate}
\item Let $\mathcal{X}_0$ be the initial domain of interest and set $k=1$;
\item Generate an appropriate design for a set of runs over the non-implausible space $\mathcal{X}_{k-1}$;
\item Identify the collection of informative outputs, $Q_k$, and obtain the corresponding simulator evaluations by running the simulator for this design of points;
\item Construct new, more accurate, emulators defined only over $\mathcal{X}_{k-1}$ for the collection $Q_k$;
\item Use the emulators to calculate implausibility across the entirety of $\mathcal{X}_{k-1}$, discarding points with high implausibility to define a smaller non-implausible region $\mathcal{X}_k$.
\item If any of our stopping criteria have been met, continue to Step 7. Otherwise, repeat the algorithm from Step 2 for wave $k+1$.
\item Generate a large number of acceptable runs from $\mathcal{X}_k$, sampled appropriately.
\end{enumerate}

By construction, it is the case that $\mathcal{X}_k \subseteq \mathcal{X}_{k-1} \subseteq \dots \subseteq \mathcal{X}_0$. Thus the history matching procedure iteratively removes parts of parameter space that are obviously unsuitable, allowing us to create emulators that are more confident in the remaining region of interest, and eventually leaving us with a well-understood, complete, non-implausible region. This approach allows us to consider only collections of outputs, $Q_k$, at each wave in a way that other methods do not: since we are removing unsuitable regions of parameter space rather than focusing on suitable regions, the absence of some outputs in a wave does not affect the validity of the points chosen. Particularly at early stages of the history matching procedure, when the parameter space can be large and contain corners of the space where the simulator behaviour is unstable, it is very useful to be able to omit outputs that cannot be satisfactorily emulated. Once we have reduced the parameter space to a point where such an output can be reliably emulated, we can reintroduce that output, with no deleterious consequences on our inference about the non-implausible space.

The stopping condition touched on in Step 6 depends on the outcome of the waves and of the eventual aim of the history match. One common stopping condition is if the emulator uncertainties are small compared to any other uncertainties in the system; if $\VAR[f_i(x)]$ is small in contrast to $\VAR[e_i]$ and $\VAR[\epsilon_{i}]$, then additional waves of emulator training will not remove any more parameter space and hence are not worth performing. Alternatively, the waves of history matching can proceed to a point where $\mathcal{X}_k = \emptyset$; this is informative, suggesting a conflict between the simulator and the real-world observation, but there is obviously no advantage to performing additional waves to `further' reduce this space. Finally, we may wish to find matches to observational data in order to reliably infer the properties of other simulated quantities, and depending on the application we may be satisfied if $M$ matches are obtained; in which case, it would be reasonable to stop the history match once we obtain the desired $M$ parameter sets (subject to considerations about the distribution of those parameter sets).

The history matching algorithm, coupled with emulation, provides a powerful toolkit for efficiently exploring the output of complex simulators in high-dimensional parameter space. The emulators allow us to predict the simulator response at points in parameter space in a fraction of the time a simulator evaluation would take, with the addendum that any such prediction comes with uncertainty; the history matching approach takes advantage of the speed of evaluation and the uncertainty in prediction to systematically, rigorously and comprehensively remove parts of parameter space that cannot result in an acceptable match to data. In so doing, this approach provides an obvious framework in which we can consider the sources of uncertainty in the simulator as well as that arising from the emulator and the observation.

One caveat appears in Step 2 of the history matching algorithm, where we specify that an `appropriate' design of currently non-implausible points is required to train the emulators. By `appropriate', we mean that the points chosen are space-filling over $\mathcal{X}_k$, thus mitigating any problems the emulators may have in determining the regression surface or the uncertainty therein \citep{santner2003design}. A main aim of the \pkg{hmer} package is to construct an accessible framework for performing the history matching and emulation procedure, and to this end ensuring that a representative sample can be elicited at each wave without recourse to user-defined approaches and the concomitant complications this brings. The function detailed in Section~\ref{sec:propose}, along with its siblings, allow a user to step through the emulation and history matching process with minimal difficulty or external specification of priors, and with no recourse to distributional assumptions that would complicate or undermine any final inference.

\section[Fundamentals]{The \pkg{hmer} package: fundamentals}\label{sec:hmer}

\pkg{hmer} is available from the Comprehensive \proglang{R} Archive Network (CRAN) at \url{https://CRAN.R-project.org/package=hmer} and can be installed and loaded in the usual manner:\footnote{The development version of the package is located on GitHub (\url{https://github.com/andy-iskauskas/hmer}); results generated here used v1.5.7.}

\begin{CodeInput}
R> install.packages("hmer")
R> library("hmer")
\end{CodeInput}

The \pkg{hmer} package, at its core, is centred around one \pkg{R6} object \code{Emulator} (itself dependent on a \code{Correlator} object) and three functions that act upon it, corresponding to emulator construction (\code{emulator\_from\_data()}), emulator validation (\code{validation\_diagnostics()}) and new point proposal (\code{generate\_new\_runs()}). We use \pkg{R6} objects for the benefits that the object-orientated approach gives, such as efficient modifying-in-place, delineation of public and private functions and arguments, and the ability to set up prototypical objects for user flexibility (a point we return to in Section~\ref{sec:proto}). We detail each of these functions in turn, its use, and the arguments provided to it. The functions have a variety of optional arguments that can be passed to it; we do not discuss all of them but highlight those that are necessary to explain the default behaviour of the function. The \pkg{hmer} documentation may be consulted for complete descriptions of the optional arguments and their usage. Descriptions of the \code{Emulator} and \code{Correlator} structure and functionality may be found in Appendix~\ref{app:emcorr}.

\subsubsection{A note on parameter sets}
In what follows, we frequently refer to collections of parameter sets. The structure of these inputs is important to bear in mind; the \code{Emulator} object and all associated functions assume that parameter sets are provided to them as a \code{data.frame} whose rows are individual parameter sets, with named columns corresponding to the names of the components of a parameter set. This is crucial for the correct determination of active variables as well as distinguishing a single $d$~dimensional parameter set from $d$ one dimensional parameter sets. Hence, for a pair of parameter sets $x_1 = (1, 1)$, $x_2 = (2, -1)$ with components \texttt{rate1}, \texttt{rate2}, a correct syntactical expression for it is

\begin{CodeInput}
R> example_points <- data.frame(rate2 = c(1, -1), rate1 = c(1, 2)) 
R> # hmer automatically arranges columns before operations
\end{CodeInput}

whereas the following will not work when passed to an \code{Emulator} function, due to the lack of named columns:

\begin{CodeInput}
R> # The below has the same issue if coerced to data.matrix or data.frame
R> points <- matrix(c(1, 1, 2, -1), nrow = 2, byrow = TRUE) 
R> # "dim(X) must have positive length"
\end{CodeInput}

\subsection[Training Emulators]{Training emulators with \code{emulator\_from\_data()}}\label{sec:emfromdata}

Often it is not practical or time-efficient to manually determine the prior specifications for an emulator, especially in situations where we wish to emulate many outputs from our simulator. Instead, we may provide the data from the simulator to \code{emulator_from_data()}, which furnishes us with a list of \code{Emulator} objects appropriately specified. The minimal syntax of \code{emulator_from_data()}, allowing it to make its best determinations about the emulator structure, is

\begin{CodeInput}
R> emulator_from_data(input_data, output_names, ranges)
\end{CodeInput}

Given a \code{data.frame} consisting of all parameter sets and their corresponding simulator outputs, the steps performed by this function for a single output $f_i(x)$ are as follows.

\begin{enumerate}
\item All input parameters are scaled to the range $[-1,1]$, using the ranges provided. This guarantees orthogonality of linear and quadratic terms in the putative basis functions: for a parameter $x_i$,
$$\langle x_i, x_i^2\rangle = \int_{-1}^{1} x_i^3 \mathrm{d}x = 0.$$
\item The regression surface is determined: basis functions in the parameters up to quadratic order (with interaction terms) form the candidate space for the collection $g_{ij}(x)$. The relevant subset is decided upon by either stepwise add or stepwise delete, as appropriate to the dimensionality of the parameter space compared to the number of simulator runs available. The final collection of terms is examined and thinned based on the proportion of variance explained by each term. The coefficients of the regression surface and their variance matrix are also collected. By default, in the presence of reasonably large numbers of simulator evaluations, we assume a separation between `global' and `local' behaviour - once determined, the regression coefficients are considered `known', so that $\VAR[\beta]=0$. This avoids identifiability issues between the regression surface and the weakly stationary surface $u_i(x)$ and can help with our physical understanding of the simulator behaviour; the assumption of a `known' regression surface is seldom problematic as the weakly stationary process $u(x)$ accounts for any and all residual variability, whether intrinsic to the output or as a consequence of the regression parameter estimation. Other options are supported: for instance, we could go to the other extreme and assume a non-informative prior for the regression coefficients, or take some compromise in between these two choices \citep{santner2003design}.
\item The basis functions determined above are used to define the collection of active variables, $x_{A_i}$. A variable is considered active if it contributes to a basis function: parameter $x_j$ is considered active for $f_i(x)$ if one of the basis functions for $f_i(x)$ is $x_j$ or $x_j^2$, or if it appears in an interaction term $x_i x_j$.
\item The residuals of the regression surface are determined from the above, and the specifications of the correlation structure are determined via a bounded maximum \textit{a posteriori} (MAP) estimate \citep{sorenson1980parameter}. Unlike in traditional maximum likelihood estimation, we do not believe in the existence of a `true' value for the hyperparameters in the correlation structure, only reasoning that the correlation length should be no larger than the greatest distance between roots of polynomial one order higher than the fitted regression surface, and no smaller than the average distance away from those same roots \citep{bower2010galaxy}. This is akin to viewing the contribution of the correlation structure as a third-order correction to the quadratic regression surface. The hyperparameter estimation includes an estimate of $\delta_i$.
\item The prior emulator for $f_i(x)$ is created using the above specifications via a call to \code{Emulator$new(...)}, and then the Bayes linear update formulae are applied to obtain the trained emulators via \code{em$adjust()}.
\end{enumerate}

For multiple outputs, this process is applied in the expected fashion: we fit regression surfaces to each output in turn, use their residuals to define the specifications for each correlation structure, then perform Bayes linear adjustment on each. This approach pointedly makes no statements about the dependency between outputs; if our univariate emulators can capture their respective outputs well, then we deem this suitable for making statements about the non-implausible space. Of course, one could impose correlations between outputs in a manner similar to that in \pkg{multivator} \citep{hankin2012multivator} -- an automated approach to this is an interesting direction for improvement within the \pkg{hmer} package.

One thing to remember is that \code{emulator_from_data()} (and its descendants, one of which we discuss in Section~\ref{sec:advanced}) will always return a named list of \code{Emulator} objects, even in the event that we wish to train to only one output. This ensures compatibility with functions in the package. A particular emulator can be accessed using the usual syntax for named lists; for example, in a list of three emulators emulating outputs $y_1$, $y_2$, $y_3$, we may access the emulator for $y_2$ via either of the following commands:

\begin{CodeInput}
R> emulators[[2]]
R> emulators$y2
\end{CodeInput}

The \code{emulator_from_data} function can be customised in almost any conceivable way: for details of the arguments that can be provided to it one may consult the associated help file. However, in the absence of any strong prior knowledge about the behaviour of the simulator outputs, the default behaviour is often appropriate to train acceptable emulators to the known data.

\subsection[Validation of Emulators]{Validation of emulators: \code{validation\_diagnostics()}}\label{sec:validation}

Having obtained emulators for our series of outputs, we must ensure that they are suitable for predicting their outputs over the current non-implausible space. As previously mentioned, a variety of diagnostic tests are possible, but the function \code{validation_diagnostics()} collects three of the most common such into one summary.

\begin{CodeInput}
R> validation_diagnostics(emulators, targets, validation, ...)
\end{CodeInput}

Here, \code{emulators} is the list of emulators to validate, \code{targets} is the set of observations that we wish to match to, and \code{validation} is a set of validation points: parameter sets for which we have simulator runs which were not provided to the emulators during training. The only mandatory argument is \code{emulators}: the omission of the argument \code{targets} precludes us from performing one of the diagnostic tests and modifies the results of the others; if \code{validation} is not provided then cross-validation is performed using the emulators' training runs (the default being leave-one-out cross-validation). For ease of notation, we refer to the set of runs upon which the emulators are tested as the `validation set', whether or not they comprise a hold-out set or whether cross-validation has been performed.

The diagnostic tests are as follows:

\begin{itemize}
\item The emulator predictions over the validation set are compared to the corresponding results from the simulator, accounting for the uncertainty. A `perfect' emulator would have $\E_{D_i}[f_i(x_j)]=f_i(x_j)$; as the emulators are statistical approximations with a well-defined uncertainty structure we instead view an emulator prediction at a point $x_i$ as satisfactory if, for a suitable choice of $c\in\mathbb{R}$,

$$f_i(x_j) \in \left[\E_{D_i}[f_i(x_j)] - c\sqrt{\VAR_{D_i}[f_i(x_j)]},\,\E_{D_i}[f(x_j)] + c\sqrt{\VAR_{D_i}[f_i(x_j)]}\right].$$

For example, $c=3$ suggests that we deem a prediction satisfactory if the true simulator evaluation lies within three standard deviations of the emulator prediction. If large numbers of points do not satisfy this requirement, it suggests a conflict between emulator and simulator, and warrants investigation - especially if the emulator is systematically over- or under-estimating the simulator output. The inclusion of the \code{targets} argument in \code{validation_diagnostics()} restricts the consideration of bad predictions to a neighbourhood of the observations, reflecting the fact that we are not overly concerned by bad diagnostics in parts of the input space far from where the simulator itself would agree with the data.

\item The emulator implausibility is compared to the `simulator implausibility', which we define in a similar vein for a parameter set $x_j$:

$$I_\text{sim}^2(x_j) = \frac{(f_i(x_j)-z_i)^2}{\VAR[e_i]+\VAR[\epsilon_{i}]}.$$

Let $\mathcal{I}_\text{em}(x_j, c)$ be the membership function for the emulator implausibility, $\mathcal{I}_\text{em}(x_j,c)=1$ if $I_i(x_j)\le c$ and $0$ otherwise, and similarly for the membership function for the simulator implausibility $\mathcal{I}_\text{sim}(x_j,c)$. Then we may summarise the options in a `classical' form\footnote{The ordering of rows and columns is chosen to match the output of the diagnostic: see, for example, Figure~\ref{fig:validation}.}:

\begin{center}
\begin{tabular}{ rr|c|c }
\multirow{2}{4em}{Simulator} & $\mathcal{I}_\text{sim}(x_j,c)=0$ & Type II & Correct \\
\cline{2-4}
& $\mathcal{I}_\text{sim}(x_j,c)=1$ & Correct & Type I \\
\hline
\multicolumn{2}{c|}{} & $\mathcal{I}_\text{em}(x_j,c)=1$ & $\mathcal{I}_\text{em}(x_j,c)=0$ \\
\multicolumn{2}{c|}{} & \multicolumn{2}{c}{Emulator}
\end{tabular}
\end{center}

For the purposes of history matching, we have no concerns about ``Type II'' errors (false positives): an emulator will often fail to rule out a point that would be ruled out by the simulator, by virtue of the additional uncertainty induced by emulation. We anticipate that later waves, when the non-implausible space has been reduced and the emulators thus trained are more accurate, will rule out any such $x_j$. However, ``Type I'' errors are of more concern: this suggests that the emulator would rule out $x_j$ when in fact the simulator output is acceptably close to the observational value. Any such point represents a failure of the emulator to accurately represent the output response across the full parameter space; there are, however, caveats. Firstly, in the context of history matching this only becomes an issue if all other emulated outputs would deem $x_j$ non-implausible: due to the structure of the emulators mismatches between emulator and simulator output can often occur in corners of space where simulator behaviour is unstable, and where the majority of observational output is far from the data. Hence a Type I error in isolation could be of concern, but the part of parameter space in which $x_j$ resides is not worthy of consideration when considering all outputs. Secondly, the usage of classical terminology is intentionally suggestive: we will see shortly that our standard implausibility cutoff is chosen under the assumption that $\sim95\%$ of the probability mass of is contained within the non-implausible region. Therefore we have an approximate Type I error rate of $\alpha = 0.05$, in accordance with normal arguments. These comments notwithstanding, examination of this diagnostic should focus on those ``Type I'' points and one should ensure that there are no systematic problems with the emulator regression surface, estimation of uncertainty, or coverage of the input parameter space. To ignore such concerns and proceed with history matching could preclude parts of parameter space that, rightfully, we should consider as being capable of giving rise to the observed reality.

Due to the fact that the calculation of implausibility requires us to have observational data $z_i$, this test cannot be run if \code{validation_diagnostics()} is not provided a \code{targets} argument.

\item The standardised prediction errors,
\begin{equation}\label{eq:standerr}
U_i(x) = \frac{f_i(x)-\E_{D_i}[f_i(x)]}{\sqrt{\VAR_{D_i}[f_i(x)]}},
\end{equation}
are calculated for each point in the validation set. Large standardised errors may suggest conflict between the emulator and simulator predictions; in most cases we may appeal to Pukelsheim once more and consider errors larger than $3$ to be of interest. At the same time, these standardised errors can indicate under-confidence or overfitting, if all of the errors are very small (i.e., less than $1$). Broadly speaking, we would expect that the errors would be unimodally distributed somewhere around $0$ with moderate extent.
\end{itemize}

The \code{validation_diagnostics()} function, by default, calculates all three of these measures for each emulated output $f_i(x)$, plotting the results of the diagnostics (and highlighting problematic points) for ease of analysis, returning a \code{data.frame} of parameter sets which failed one or more diagnostics. We will see this in action in Section~\ref{sec:ex}.

\subsection[Proposing New Points]{Proposal of new points using \code{generate\_new\_design()}}\label{sec:propose}

Suppose we have trained a set of emulators, and by recourse to their diagnostics are satisfied that they can sufficiently emulate their outputs over the parameter region of interest. We now wish to use these emulators to determine the new non-implausible region. As mentioned in Section~\ref{sec:hm}, to successfully train a set of emulators for the next wave we require an `appropriate' design over the non-implausible space: ideally it should be space-filling and uniform across the space. For most (if not all) applications, a direct parametrisation of the non-implausible space is impossible, so we cannot simply define a membership function for the non-implausible space $\mathcal{X}_{k}$ at wave $k$. Instead, we use a variety of methods to try to ensure that we understand the boundary of $\mathcal{X}_k$ as well as suitably populate the interior of the space.

The minimal specification for the function \code{generate_new_design()} is

\begin{CodeInput}
R> generate_new_design(emulators, n_points, targets)
\end{CodeInput}

The \code{emulators} and \code{targets} arguments are the same as that of \code{validation_diagnostics()}; the \code{n_points} argument simply indicates how many points we wish to generate from the non-implausible space $\mathcal{X}$. The normal behaviour of the function employs the following techniques:

\begin{enumerate}
\item A large Latin hypercube design (LHD) \citep{mckay2000comparison} is generated across the minimum enclosing hyperrectangle (MEH) of the non-implausible space $\mathcal{X}$. The points in the design are rejected if their implausibility exceeds our cut-off.

\item The points that remain are randomly pairwise-selected (with greater weight given to those points which have larger separation) and points are sampled along a ray containing the pair of points which extends to the edge of the MEH. Implausibilities along the ray are calculated, with points being retained if they are non-implausible and satisfy one of the following:
\begin{enumerate}
    \item An immediate neighbour of the point on the ray is deemed implausible;
    \item The point is one of the terminating points on the ray; that is, it lies on the boundary of the space.
\end{enumerate}

\item The remaining points from the LHD, supplemented by the boundary points, are used to define the centres for a collection of ellipsoids, from which we may generate a mixture distribution of uniform ellipsoids. The radii of the ellipsoids are such that each ellipsoid extends to the boundary of the MEH, and so in general ellipsoids overlap significantly - optimal radii for the ellipsoids are determined using a burn-in phase. When proposing points, an importance sampling argument is used to avoid over-sampling at the intersection of ellipsoids.

\item The space is resampled: the complete set of proposal points are thinned to half of the number desired, according to a maximin argument. Steps 2 and 3 are performed again.

\item We apply thinning once more to obtain \code{n_points} points with a maximin argument.
\end{enumerate}

\begin{figure}[h!]
\centering
\begin{subfigure}{0.3\textwidth}
\includegraphics[width=\textwidth]{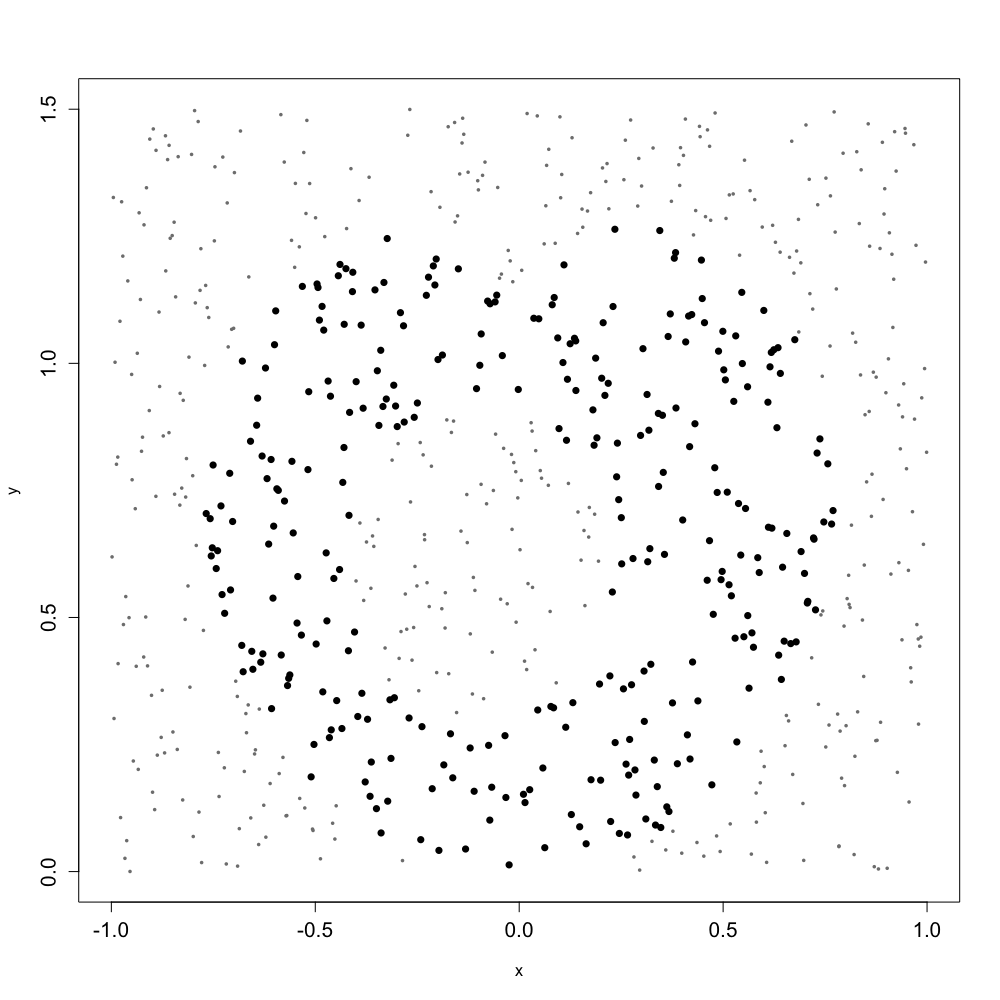}
\end{subfigure}
\begin{subfigure}{0.3\textwidth}
\includegraphics[width=\textwidth]{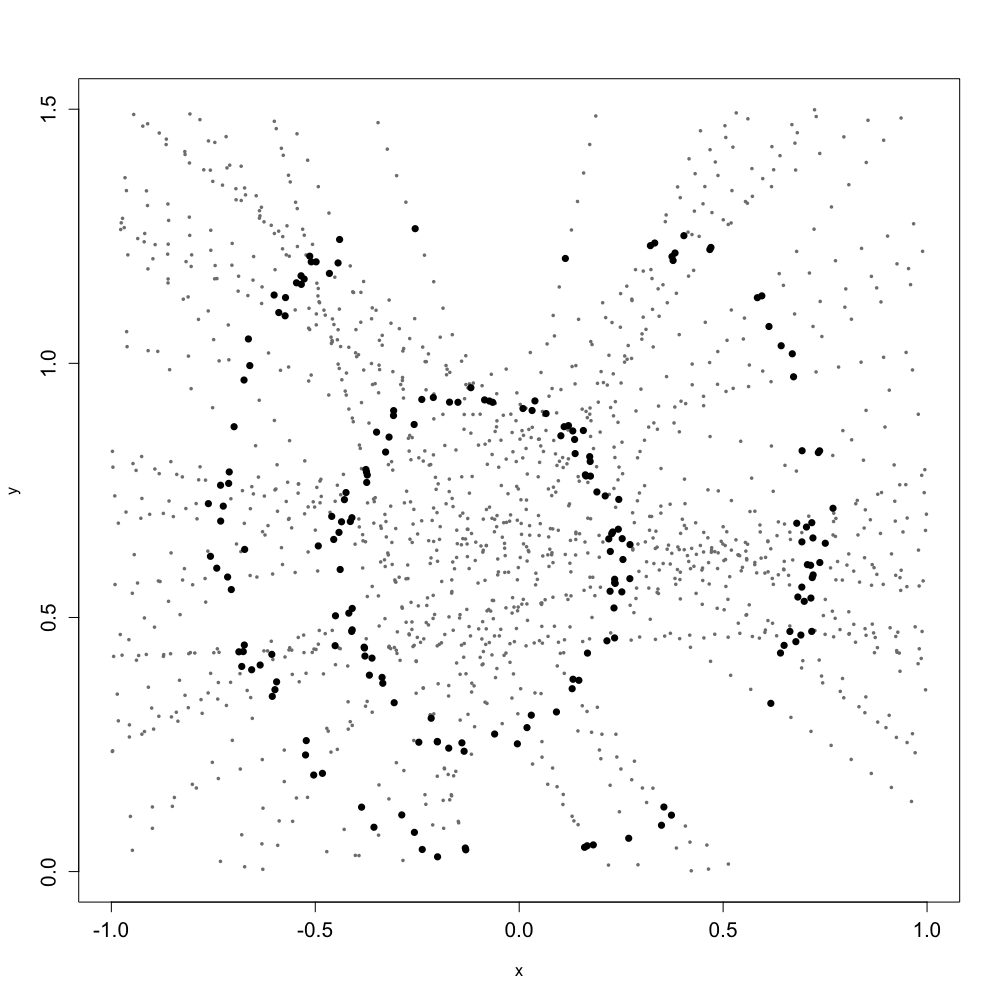}
\end{subfigure}
\\
\begin{subfigure}{0.3\textwidth}
\includegraphics[width=\textwidth]{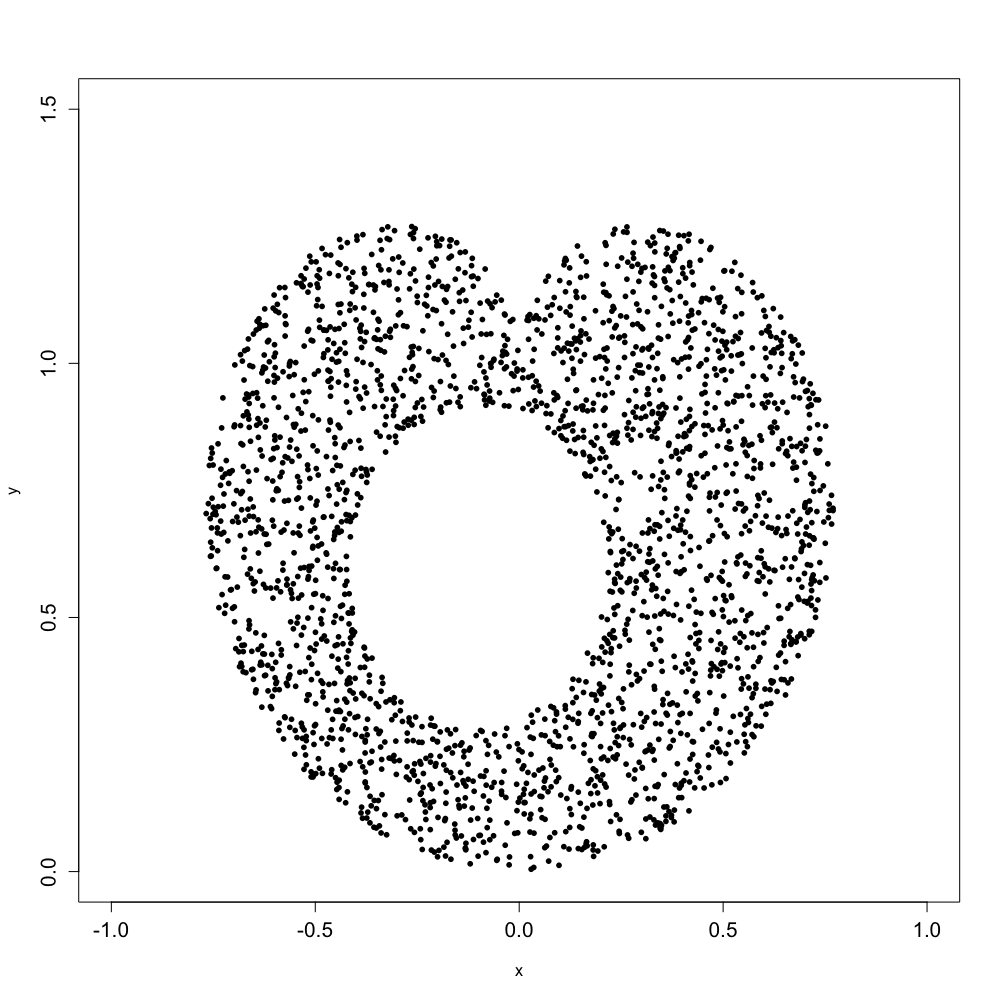}
\end{subfigure}
\begin{subfigure}{0.3\textwidth}
\includegraphics[width=\textwidth]{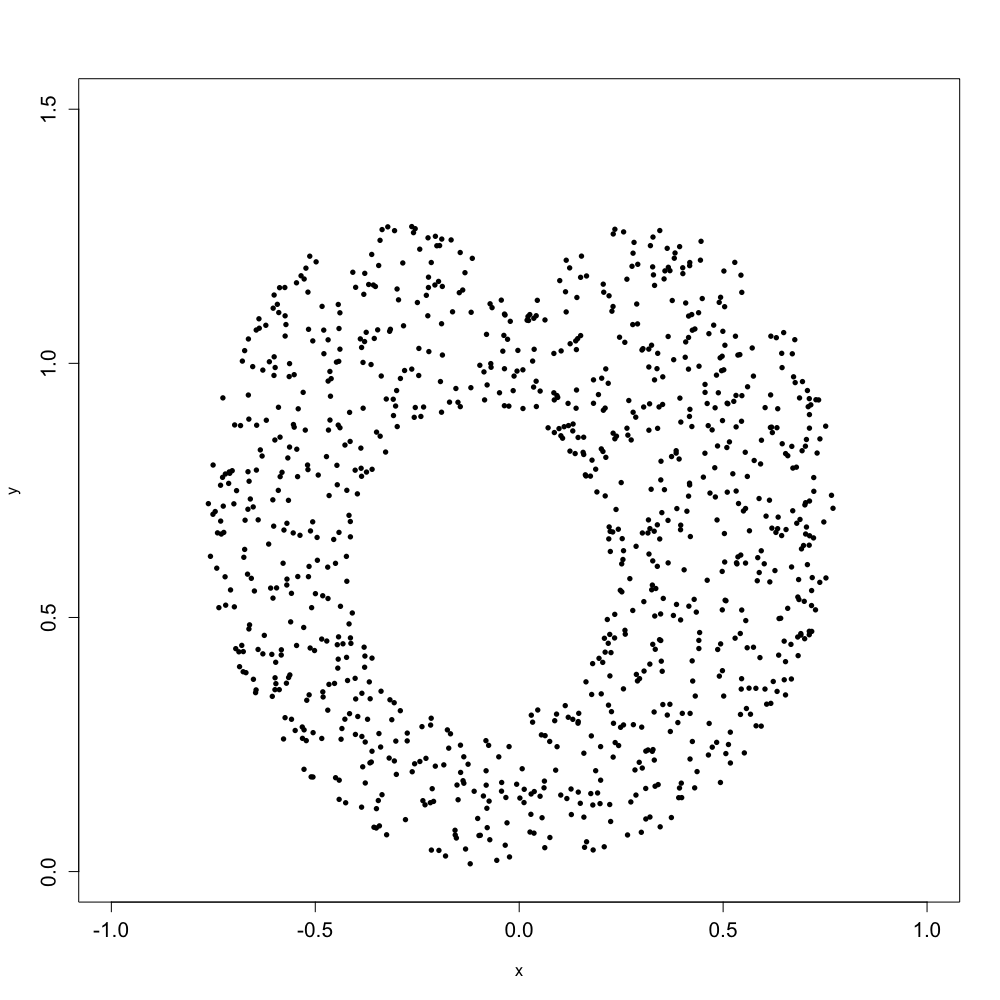}
\end{subfigure}
\caption[Demonstration of sampling]{\small{From top-left to bottom-right, the steps in \code{generate\_new\_runs}, applied to an artificial example of a non-implausible space: Latin hypercube sampling with rejection; line sampling, uniform ellipsoid importance sampling, and a final thinning stage. Points in grey are rejected points (in the case of Latin hypercube sampling and line sampling) or the initial set of points defining the mixture distribution (for importance sampling).}}
\label{fig:propdemo}
\end{figure}

We can see a diagrammatic demonstration of these steps in Figure~\ref{fig:propdemo}, excepting Step 4 (the resampling step). For this, a contrived example of a non-implausible space was defined: the space is a cardioid with a hole removed from inside it. We can see that the line sampling step (Step 2) picks up both the external and the internal boundaries, as we would expect.

While Figure~\ref{fig:propdemo} demonstrates the process for a simple 2~dimensional problem, the challenge of uniformly exploring a small space in higher dimensions is difficult. For a high-dimensional problem, the number of corners in the space increases exponentially, and it can be infeasible to investigate all of the corners (let alone the complete boundary). However, missing such corners is not disastrous in the context of history matching; parts of the boundary that are missed will continue to have high emulator uncertainty and therefore not be ruled out until we sample from that region. For specific problems where we have some intuition for or expectation of viable parameter combinations, there may be other approaches that would give a more representative sample of the space; we discuss briefly how these could be incorporated in Section~\ref{sec:advanced}.

The default approach is designed to leverage the emulators' comparative speed in evaluating implausibility -- an emulator evaluation takes $\mathcal{O}(10^{-5})$ seconds which vastly outpaces most simulator outputs -- in order to explore both the boundary and the interior of the non-implausible region in a straightforward manner. With this in mind, the steps above can efficiently provide a uniform sample from a non-implausible space, while also giving us adequate coverage on the boundary of the space. Were we to train emulators on the proposed points, the boundary points therein will aid in the learning of the regression terms while the points in the interior will aid in the learning of the weakly stationary process.

\section{Example: Application to an SIRS model}\label{sec:ex}

We now use the core functions and objects described in Section~\ref{sec:hmer} on a simple toy SIRS epidemic model, representing movement through a system where an individual can be (S)usceptible to the infection, (I)nfected with it, or (R)ecovered from it. Recovered individuals can return to the Susceptible state, representing a loss of immunity over time. While simple, it will suffice to demonstrate the functionality we wish to show. The model in question can be described by the following differential equations:

\begin{align*}
\frac{\ud S}{\ud t} &= \alpha_{SR}R - \frac{\alpha_{SI}SI}{S+I+R}, \\
\frac{\ud I}{\ud t} &= \frac{\alpha_{SI}SI}{S+I+R}-\alpha_{IR}I,\\
\frac{\ud R}{\ud t} &= \alpha_{IR}I-\alpha_{SR}R.
\end{align*}

The model has three states $S$, $I$, and $R$ representing susceptible, infected, and recovered individuals; three input parameters, $\alpha_{SI}$, $\alpha_{IR}$ and $\alpha_{SR}$ representing rate of infection, recovery, and waning immunity respectively; and we will focus on three outputs $nS\equiv S(10)$, $nI\equiv I(10)$, $nR \equiv R(10)$, namely the numbers of susceptible, infected, and recovered individuals at time $t=10$. The choices of a single output from each compartment are purely for demonstrative purposes; we choose to investigate these outputs at $t=10$ in order to allow sufficiently interesting behaviour across the parameter space without reverting to complete die-out of the disease modelled. We also note that, in the absence of vital dynamics, we would expect $S(t)+I(t)+R(t)=N$ constant, which provides a good check on the performance of the emulators (which are not, \emph{a priori}, provided with this constraint). The model is initialised at $t=0$ with $S(0)=950$, $I(0)=50$ and $R(0)=0$. The fact that this model is low-dimensional and consists of relatively uncomplicated differential equations means that we could generate a vast number of simulator evaluations and forgo emulation entirely: this is seldom the case in real-world applications. However, even in this simple example the emulator evaluation time is faster than that for the simulator by two orders of magnitude; for more complex simulators and applications, this difference becomes even more pronounced and forms a large part of the motivation for employing emulation.

The model considered is comparatively simple, for ease of evaluation, while being complex enough to necessitate functionality of the package that we wish to detail below. Examples of more complex applications of emulation and history matching, including some using \pkg{hmer} directly, can be found in \cite{andrianakis2015bayesian, craig1997pressure, bower2010galaxy, scarponi2023tbhiv, vernon2022bayesianjune}.

\subsection{Initial simulator runs and emulator training}\label{sec:initmodel}

The code required to run this example, and generate the plots therein, is included in the Supplementary Material to this paper. To set up the first wave of history matching, we require a set of simulator runs across the full parameter space in question. One option would be to create a Latin hypercube design with the desired number of points before putting them into the simulator function: sample functions for doing so can be found at the top of the Supplementary Material (namely the functions \code{ode_results()}, \code{get_res()}, and \code{get_lhs_design()}). Methods that allow us to cover the space and encompass as much of the interesting simulator behaviour as possible are reasonable for producing a first design \footnote{However, the \pkg{hmer} package has a collection of pre-built datasets which includes the requisite runs from this simulator and parameter space: they can be accessed using \code{data("SIRSample")}, and we will use these henceforth.} \citep{santner2003design} .

The ranges of our parameters are $\alpha_{SI}\in[0.1, 0.8]$, $\alpha_{IR}\in[0, 0.5]$, $\alpha_{SR}\in[0, 0.05]$. This choice of ranges is somewhat artificial, but our general principle is that we choose initial ranges to be large enough to cover all feasible parameter combinations without being unphysical. Wider ranges will, of course, result in more `waves' of history matching: balancing extensive ranges against computational demands is generally a judgement to be determined by the model expert. The `observations' we wish to match to are synthetic, as this model has no real world analogue; to these targets we have added some uncertainty reflecting the range of observation we would see in reality. Both ranges and targets are defined as follows.

\begin{CodeInput}
R> ranges <- list(aSI = c(0.1, 0.8), aIR = c(0, 0.5), aSR = c(0, 0.05))
R> targets <- list(nS = c(580, 651),
                   nI = list(val = 169, sigma = 8.45),
                   nR = c(199, 221))
\end{CodeInput}

Given the synthetic nature of the model we obtained targets by running the equivalent stochastic model at a given parameter set, using this as a proxy for observation of the quantities. In concrete applications such targets arise from observational studies or secondary modelling (for example, data on confirmed cases of a disease and the derived prevalance or incidence based on observation). Note that we have two different conventions for specifying targets and their uncertainties. The one which most aligns with the mathematical discussion of observation error is that for the output \code{nI}, where we observe a value with a corresponding uncertainty. Such a framework is not always possible; for distributional or other reasons we may only be able to state that we know that our target value lies within a specified interval. These two conventions can be used concurrently in a set of targets as deemed necessary. 

\begin{figure}[!h]
\centering
\includegraphics[width=0.55\textwidth]{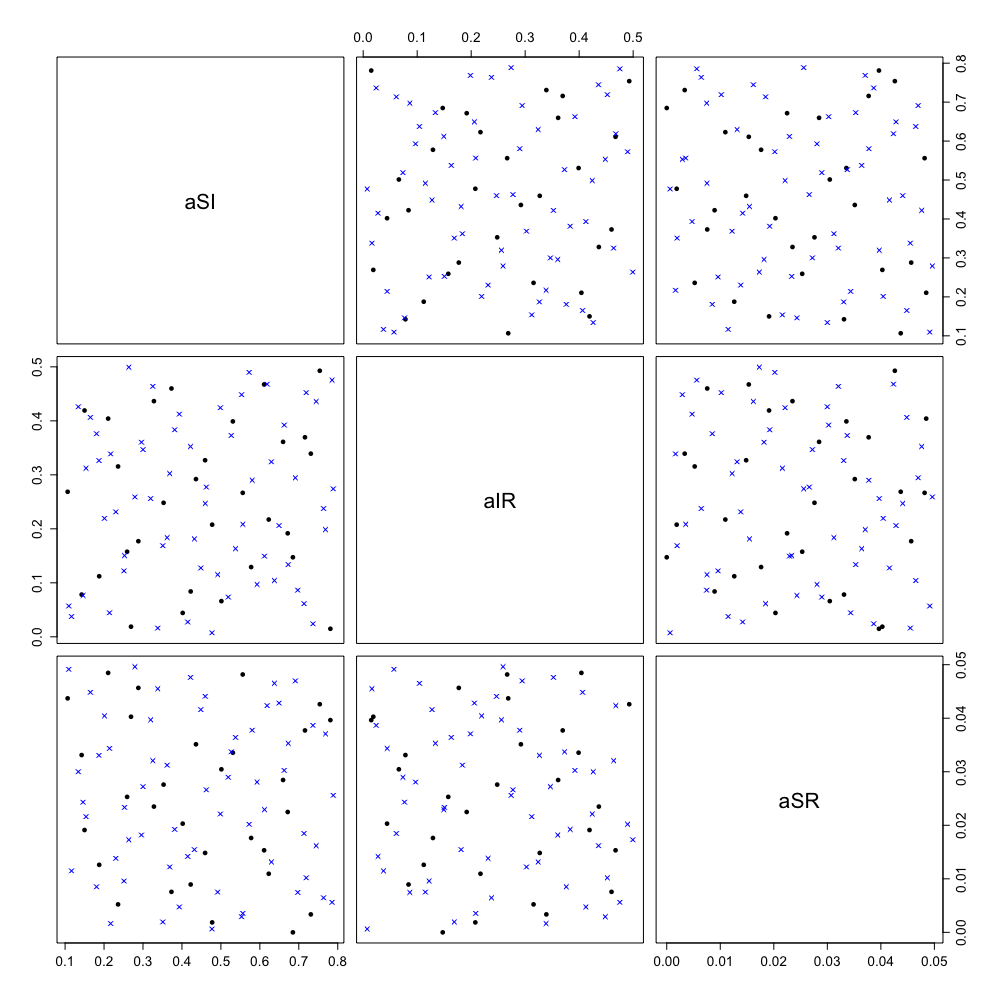}
\caption{The initial set of simulator runs: runs that we will use for training are plotted as circles and those for validation as crosses.}
\label{fig:initiallhs}
\end{figure}

Our initial simulator runs, consisting of $30$ training runs and $60$ validation runs, are shown in Figure~\ref{fig:initiallhs}. Note that we plan to use only $30$ runs to train the emulators over the entire space: this is the minimum that we should consider using for a $3$~dimensional system such as this\footnote{As a general rule of thumb, one should aim to have at least $10d$ points for a $d$~dimensional parameter space as training runs \citep{loeppky2009choosing}.}. Nevertheless, this is very little information required by the emulators for us to make inferences about the structure of the entire space.

The choice of the size of the training and validation sets, as well as the spread of points within, is a model-dependent decision: generally we would try to ensure both sets are representative of model behaviour, in that they cover the input parameter space insofar as is possible. The size of the validation set may vary depending on the computational cost of obtaining simulator evaluations, and depends on expert judgement and understanding of model constraints. With the initial parameters and data set up, we use \pkg{hmer} to train some emulators to the three targets.

\begin{CodeChunk}
\begin{CodeInput}
R> ems_wave1 <- emulator_from_data(SIRSample$training, names(targets), ranges)
\end{CodeInput}
\begin{CodeOutput}
[1] "Fitting regression surfaces..."
[1] "Building correlation structures.."
[1] "Creating emulators..."
[1] "Performing Bayes linear adjustment..."
\end{CodeOutput}
\end{CodeChunk}

There are a number of messages that indicate our progress through emulator training; in the event where there are many more outputs to train to, further messages are displayed to indicate which output is being considered. Once trained, we may examine the structure of an emulator by invoking its built-in \code{print()} statement: for example, to view the emulator for the $nI$ output, we can enter the following.

\begin{CodeChunk}
\begin{CodeInput}
R> ems_wave1$nI
\end{CodeInput}
\begin{CodeOutput}
Parameters and ranges:  aSI: c(0.1, 0.8): aIR: c(0, 0.5): aSR: c(0, 0.05) 
Specifications: 
	 Basis functions:  (Intercept); aSI; aIR; I(aIR^2); aSI:aIR 
	 Active variables aSI; aIR 
	 Regression Surface Expectation:  149.8096; 199.5466; -281.9466;
	  201.6298; -196.1621 
	 Regression surface Variance (eigenvalues):  0; 0; 0; 0; 0 
Correlation Structure: 
Bayes-adjusted emulator - prior specifications listed. 
	 Variance (Representative):  3226.426
	 Expectation:  0 
	 Correlation type: exp_sq 
	 Hyperparameters:  theta: 0.9033
	 Nugget term: 0.05 
Mixed covariance:  0 0 0 0 0
\end{CodeOutput}
\end{CodeChunk}

The \code{print} statement details the basis functions chosen, the corresponding regression coefficient prior specifications, and the active variables, as well as the specifics of the \code{Correlator} object: particularly its prior variance $\sigma^2$, hyperparameters for the chosen correlation function, and the nugget term $\delta$. Note that the \code{Emulator} print statement also informs us that the emulator has been Bayes-adjusted. We can access the unadjusted emulator, should we wish to (for example to make adjustments to the correlation structure or regression surface), by calling \code{ems_wave1$nI$o_em}. We may already note from this printout that the variable $\alpha_{SR}$ is not anticipated to be informative for the number of infected people at $t=10$, due to its absence in the active variable set. This is a partial justification for our choice of a `known' regression surface - simply by examining the structure of the regression surface, we can determine physically interesting properties of the simulator. Here, the constructed emulator lends weight to a qualitative observation about the system: we would not expect waning immunity to have had a substantial effect on the output of the simulator at $t=10$, since it is unlikely that we have had a large number of individuals infected, recovered, returned to susceptibility, and infected once more. Such observations can be extremely useful in motivating further explorations of particular regions of the parameter space of the simulator, and we may obtain these insights through a simple print statement in the package.

We, in principle, now have a collection of emulators for the outputs of interest, and can therefore explore the parameter space without recourse to simulator evaluations.

\subsection{Validating the emulators}

\begin{figure}[!h]
\centering
\includegraphics[width=0.8\textwidth]{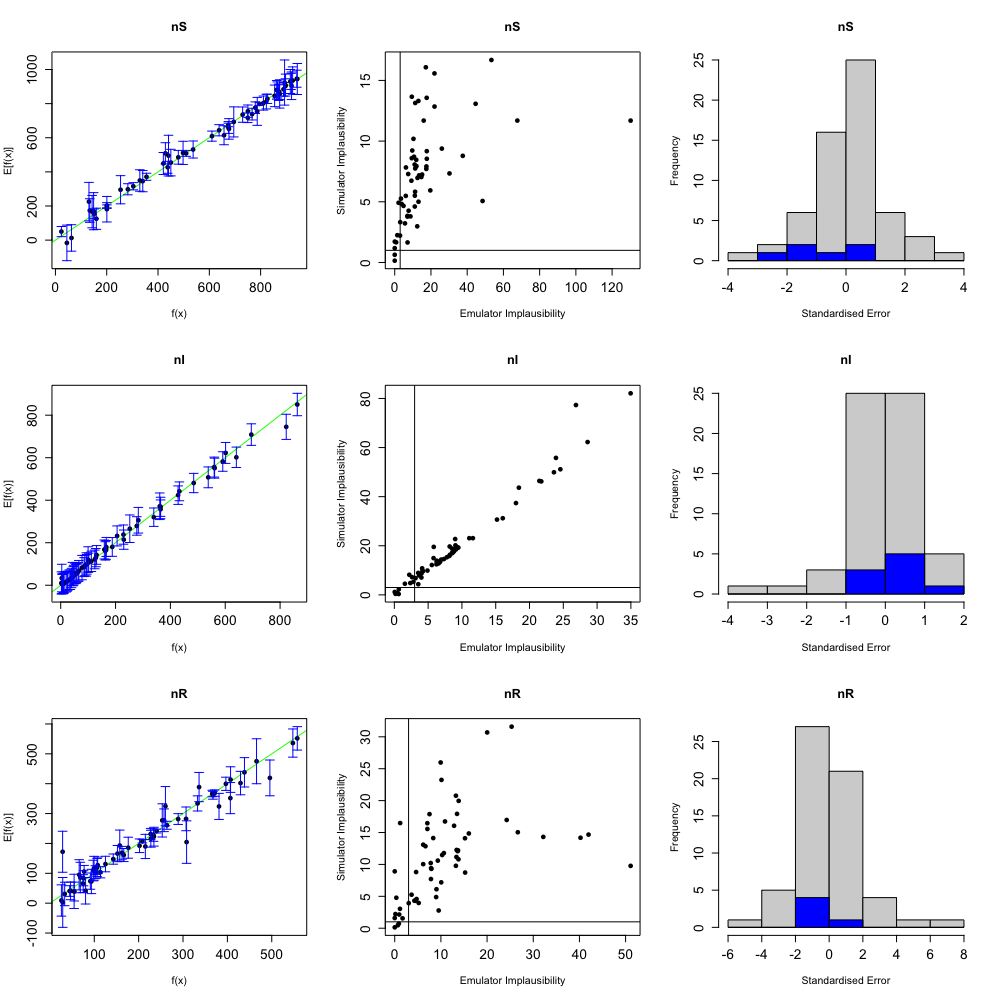}
\caption{Plots of the validation diagnostics for the SIR emulators.}
\label{fig:validation}
\end{figure}

Before we do anything else with these emulators, we must ensure they are suitable for inference. We use the \code{validation_diagnostics()} function to check the quality of their predictions with respect to unseen simulator runs. Along with the following code output, it produces a plot of the diagnostics which we include in Figure~\ref{fig:validation}.

\begin{CodeChunk}
\begin{CodeInput}
R> validation_diagnostics(ems_wave1, targets, SIRSample$validation)
\end{CodeInput}
\begin{CodeOutput}
[1] aSI aIR aSR
<0 rows> (or 0-length row.names)
\end{CodeOutput}
\end{CodeChunk}

The output of \code{validation_diagnostics()} is an empty \code{data.frame}, indicating no conflict between the emulators and the simulator outputs. Each row of the plot corresponds to an emulated output; each column to a different diagnostic. 

\begin{figure}[h]
\centering
\begin{subfigure}{0.45\textwidth}
\includegraphics[width = \textwidth]{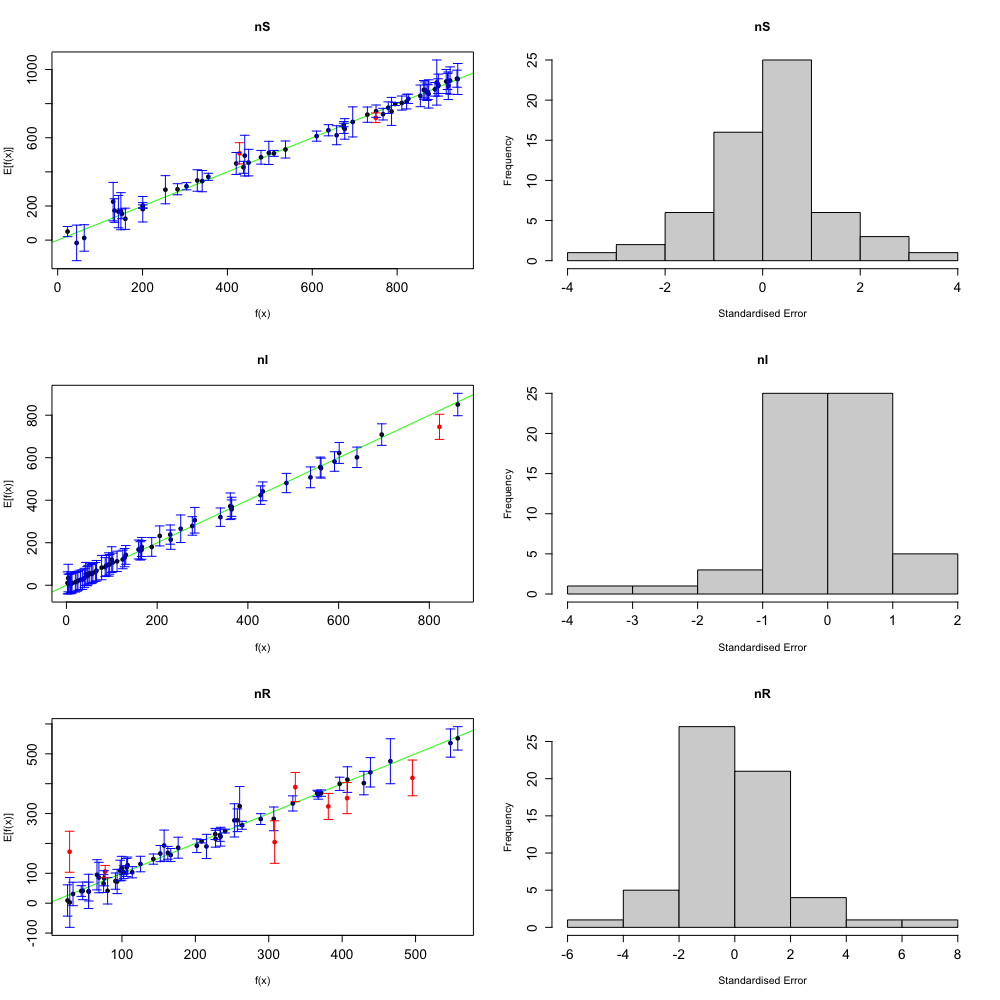}
\end{subfigure}
\begin{subfigure}{0.45\textwidth}
\includegraphics[width = \textwidth]{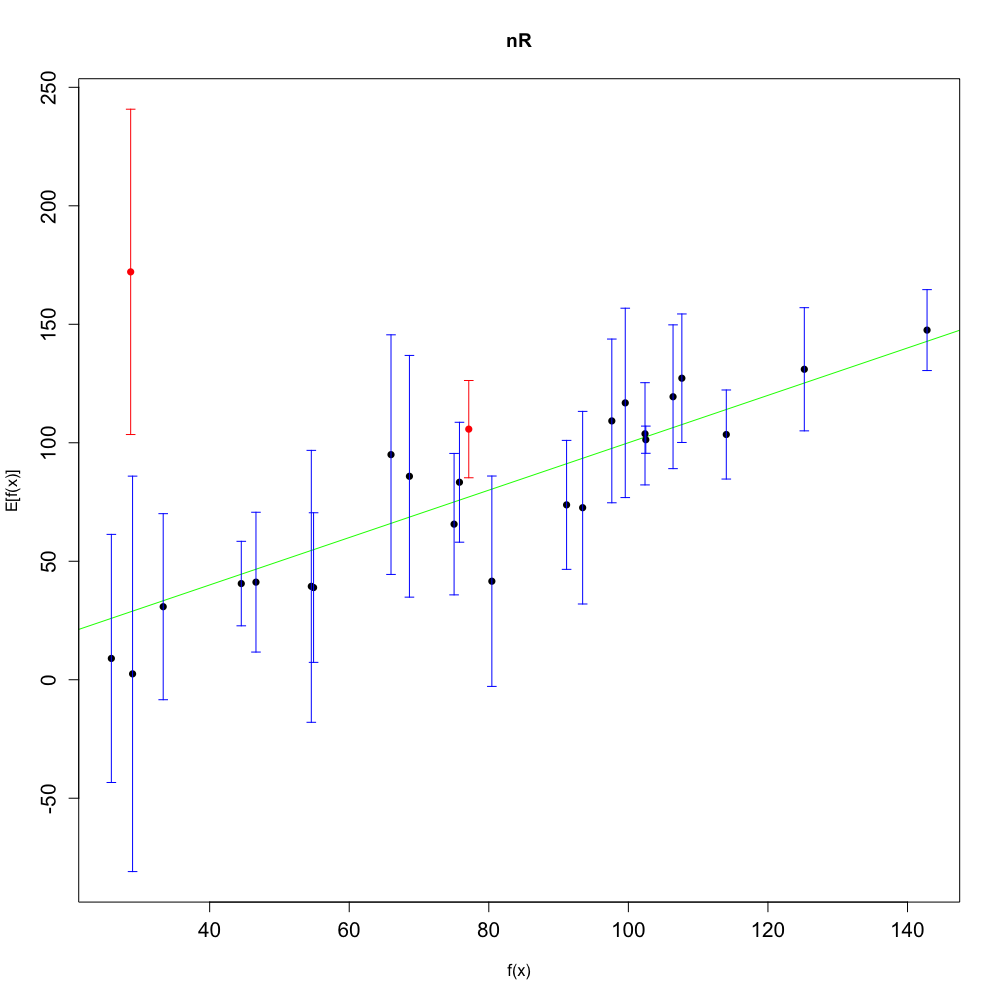}
\end{subfigure}
\caption[Validation Diagnostics: No Targets]{Validation diagnostics without provided targets for all emulators (left) and comparison diagnostics for $n_R$ (right).}
\label{fig:validationnotarg}
\end{figure}

The first column compares the simulator output ($x$-axis) with the emulator prediction ($y$-axis): the green line represents the line of `perfect' fit where $f_i(x)=\E_{D_i}[f_i(x)]$. The error-bars encode the emulator uncertainty around its prediction. We can see that almost all error-bars contain the green line. There are a few exceptions to this, where error-bars just miss the line of perfect fits --- in particular, the prediction in the lower range of the \code{nR} emulator (bottom-left plot). The reason these have not been judged to have failed diagnostics is due to the fact that the simulator predictions are a long way away from the observation --- according to the simulator the parameter set in the bottom of the \code{nR} plot gives a value in the region of $10$ recovered people, when in actual fact we anticipate any relevant points to provide a value closer to $200$ as a minimum. If we omit the \code{targets} argument from the \code{validation_diagnostics()} function call, these points will be highlighted in red, and the function would return a non-empty \code{data.frame}, as shown in Figure~\ref{fig:validationnotarg} in general and for the \code{nR} emulator specifically.

The second column considers potential emulator misclassification. On the $x$-axis is emulator implausibility for each point; on the $y$-axis is simulator `implausibility' as defined in Section~\ref{sec:validation}. A vertical and horizontal line are placed on the plot to indicate the relevant cut-off values for each of the simulator and emulator. Points in the upper-right or lower-left segments fall into the first category detailed in the relevant part of Section~\ref{sec:validation} (simulator agrees with emulator) and points in the upper-left segment fall into our second category (emulator does not rule a point out); neither of these are cause for concern. A point in the lower-right quadrant suggests that it would be ruled out as implausible by the emulator but not by the simulator, would be highlighted in red, and would warrant further investigation as already discussed in Section~\ref{sec:validation}.

The final column provides a histogram of the standardised errors \eqref{eq:standerr}. Here, we are simply looking to see a reasonable spread centred around $0$, with no more than a couple of outliers (points for which $U_i(x)$ has magnitude greater than around $3$, say). We see that this is the case; there is one potentially egregious point in the bottom-right plot, but we expect this due to the considerations of the corresponding plot in the first column. %This is reinforced by the distribution of parameter sets whose simulator output would match the observational data, here highlighted in blue; none of these points have high standardised error.

A number of other diagnostic functions are available in the package, including a series of plots obtained from \code{individual_errors()} which implements many of the tests described in \cite{bastos2009diagnostics}, \code{residual_diag()} which allows us to look directly at the performance of the regression surface, and distributional tests for goodness-of-fit within \code{summary_diag()}. Each of these functions takes a single emulator as an argument, and these can be used to supplement any intuition gained from \code{validation_diagnostics()}.

\subsection{Emulator and History Matching visualisation}

The major advantage of emulation is that we can efficiently explore the parameter space without a large number of (usually computationally expensive) simulator runs. Here we detail a few functions in \pkg{hmer} that allow us to inspect and gain insight into the simulator behaviour across the full parameter space, not just those parts for which we have runs.

The most useful function for this purpose is \code{emulator_plot(emulators, plot_type, ...)} which produces a contour plot of a desired emulator statistic over a two-dimensional slice of the parameter space. It can be called with a collection of emulators as its first argument, or directly from a single emulator via the plot command. Either option returns a \pkg{ggplot2} object which, by the nature of such objects, can be modified after the fact to add other plot objects to it\footnote{The first of these commands returns a \code{ggmatrix} of plots, so augmentation of individual plots is more involved but still possible. For details on modifying these objects, consult the \pkg{ggplot2} \citep{wickham2021ggplot} and \pkg{GGally} \citep{schloerke2021ggally} packages.}.

\begin{CodeInput}
R> emulator_plot(ems_wave1)
R> ems_wave1$nR$plot(plot_type = 'sd', params = c('aSI', 'aIR'),
       fixed_vals = c(aSR = 0.045)) +
       geom_point(data = SIRSample$training, aes(x = aSI, y = aIR))
\end{CodeInput}

\begin{figure}[h]
\centering
\includegraphics[width = 0.7\textwidth]{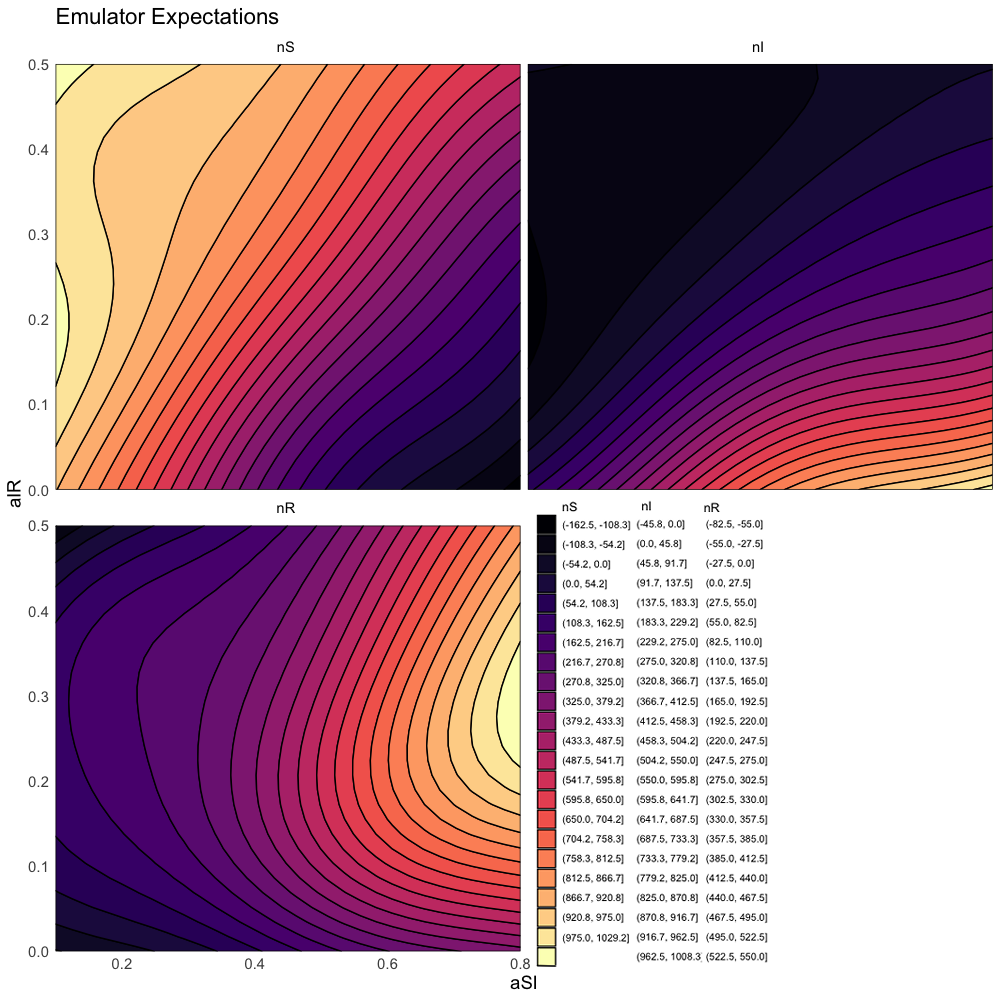}
\caption[Emulator Plot Output]{The output from the \code{emulator\_plot} function. Each plot displays the qualitative behaviour of the output across the particular two-dimensional slice of parameter space.}
\label{fig:emplot}
\end{figure}

The output of the first of these commands is shown in Figure~\ref{fig:emplot}. For multiple output plots, no scale is typically shown as we will likely be considering trends of behaviour, and individually plot those emulators where magnitude of prediction is paramount: here, we have included them for completeness. We can see the perturbations arising from the Bayes linear update beyond that of the regression surface: recall that the regression surfaces consist of constant, linear and quadratic terms, which would not be sufficient to describe the contour lines we see here.

Indeed, one can view the unadjusted emulators using a very similar plotting argument so as to make a direct comparison by using the \code{em$o_em} argument: for example to examine the prior (untrained) emulator for output $nR$

\begin{CodeInput}
R> plot(ems_wave1$nR$o_em)
\end{CodeInput}

\begin{figure}[h]
\centering
\begin{subfigure}{0.45\textwidth}
\includegraphics[width = \textwidth]{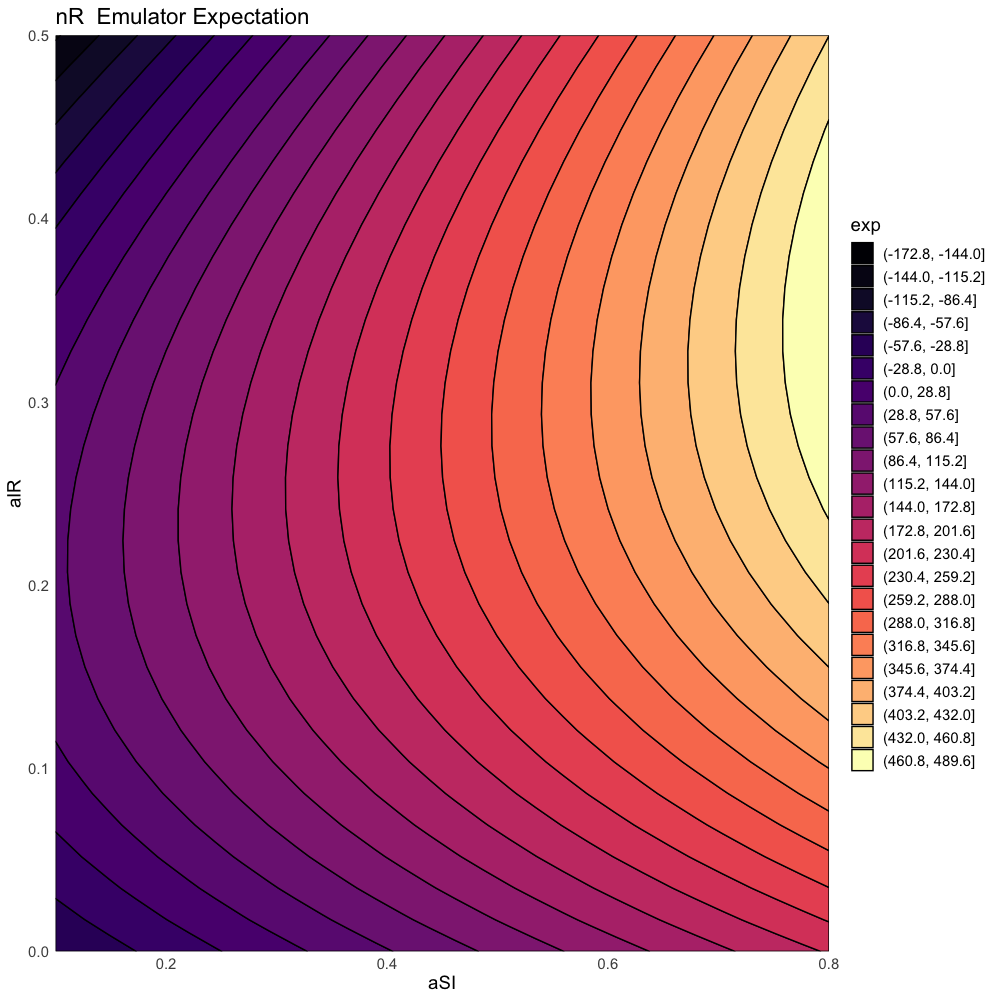}
\end{subfigure}
\begin{subfigure}{0.45\textwidth}
\includegraphics[width = \textwidth]{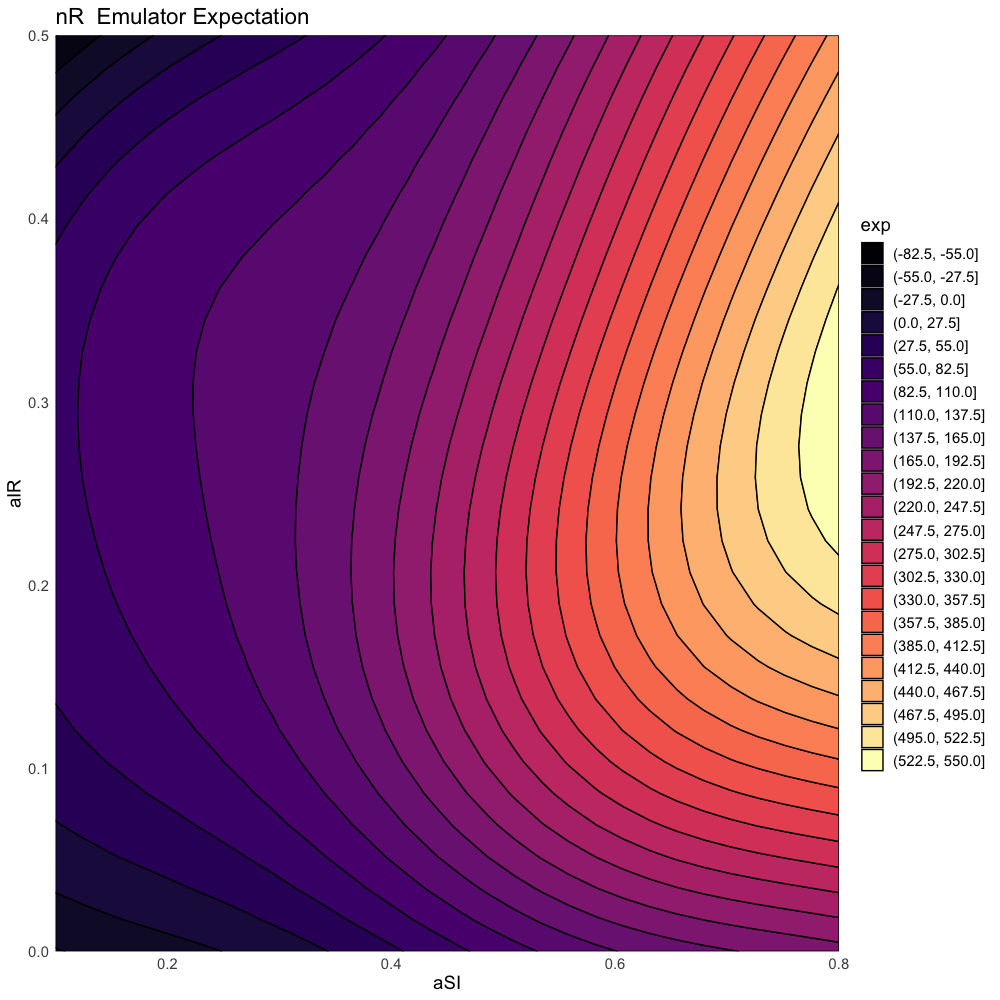}
\end{subfigure}
\caption[Emulator Comparison]{The emulator expectation for output $nS$ before (left) and after (right) Bayes linear adjustment.}
\label{fig:embeforeafter}
\end{figure}

The result of this is shown in Figure~\ref{fig:embeforeafter}. We have plotted the unadjusted and the adjusted emulators on their own in order to include the scales over which they are defined, and can see clearly the effect of the training points on the structure of the output surface as predicted by the emulator.

The \code{emulator_plot()} command takes a variety of arguments alongside the emulators. Some have been used in the examples above: the \code{params} argument determines which two parameters are to be plotted with respect to, and the \code{fixed_vals} argument determines the fixed values of the remaining parameters. The \code{plot_type} command can be one of five options: \code{"exp"} for emulator expectation (the default); \code{"var"} for emulator variance; \code{"sd"} for emulator standard deviation, \code{"imp"} for emulator implausibility; and \code{"nimp"} for $n$th maximum implausibility. The latter two arguments require \code{targets} to be specified; the final argument can also accept an \code{nth} argument to determine which maximised implausibility to plot. We may also increase the fidelity of the plot using the points-per-dimension (\code{ppd}) argument.

\begin{figure}[!h]
\centering
\begin{subfigure}{0.45\textwidth}
\includegraphics[width = \textwidth]{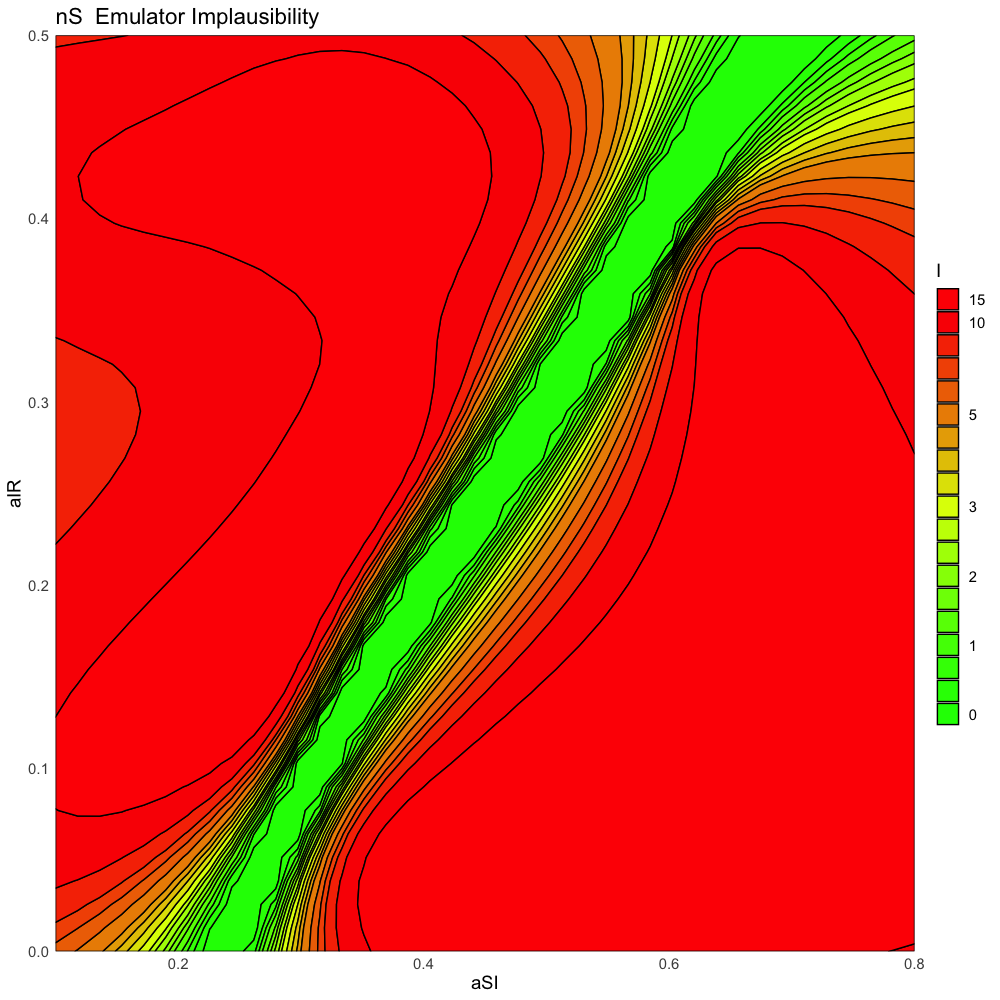}
\end{subfigure}
\begin{subfigure}{0.45\textwidth}
\includegraphics[width = \textwidth]{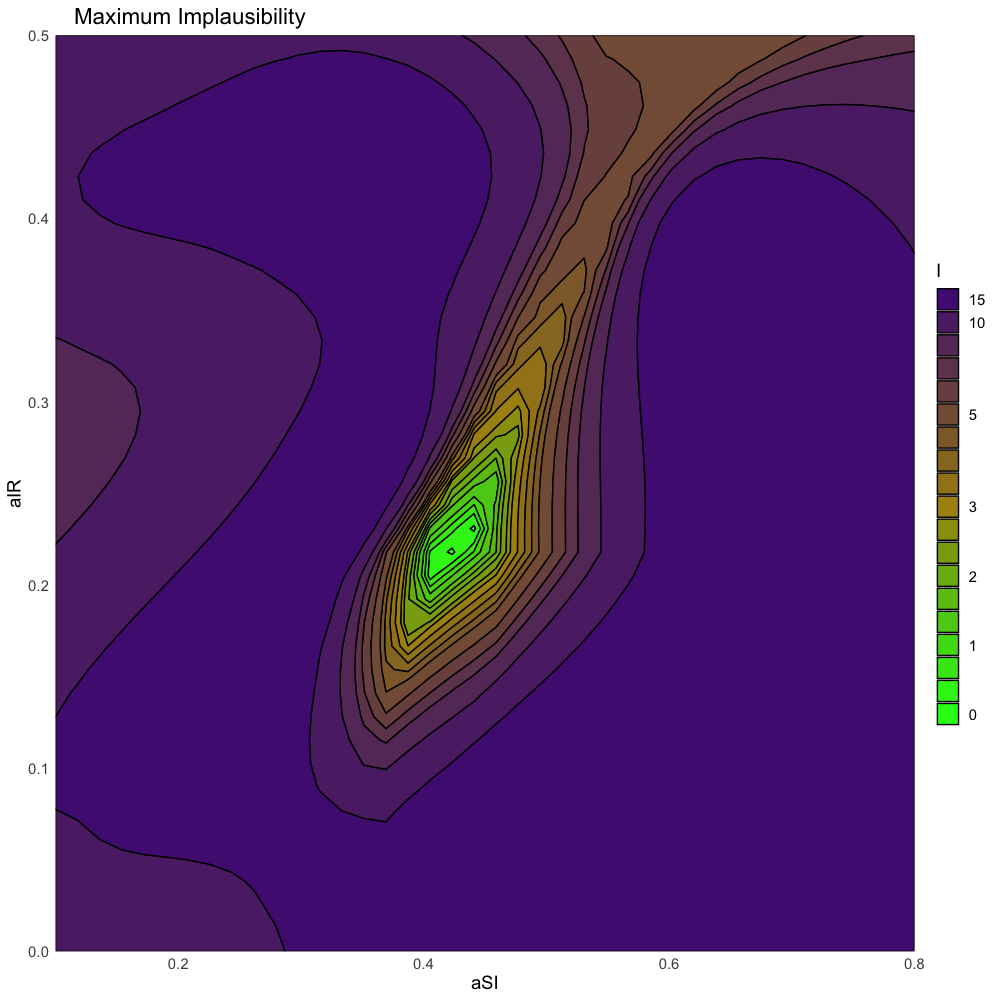}
\end{subfigure}
\caption[Implausibility Plots]{The emulator implausibility for a single emulated output (in this example, the number of susceptible people, $nS$) and the maximum implausibility over all three emulated outputs.}
\label{fig:impplots}
\end{figure}

Central to the history matching procedure is the definition of an implausibility measure, as the choice of a cutoff for a point to be deemed non-implausible is integral to the extent to which the parameter space is reduced on subsequent waves. Correspondingly, the plots from \code{emulator_plot()} where we visualise implausibility of an emulated output (via \code{plot_type = `imp'}) or implausibility of a collection of emulators as considered in Section~\ref{sec:hm} (via \code{plot_type = `nimp'}) can be critical in making determinations on what constitutes an acceptable cutoff. We will discuss other means of selecting an appropriate implausibility cutoff shortly, but we first note that we may use the emulator plots to examine the non-implausible space under a number of putative cutoffs, as well as the geometric structure of the space. The output of each of these commands is shown in Figure~\ref{fig:impplots}, and gives an indication of the contribution of the $nS$ emulator (left) to our overall measure (right): a clear correlation between acceptable values of \code{aSI} and \code{aIR} is present if we wish to simply match to the number of susceptible people, but the structure of the overall space when all outputs are considered limits this simple relationship considerably. After considering the shapes and size of the contours, were we to determine that the amount of space that would be reduced at a given implausibility cutoff was too severe or too conservative we may apply this knowledge in the history matching procedure by selecting an informed value for the cutoff.

\begin{CodeInput}
R> emulator_plot(ems_wave1$nS, plot_type = "imp", targets = targets)
R> emulator_plot(ems_wave1, plot_type = "nimp", targets = targets,
      ppd = 40, cb = TRUE)
\end{CodeInput}

\noindent\fbox{%
    \parbox{\textwidth}{%
	\textbf{Implausibility Plots and Colourblindness}
	
In any plots which calculate and display implausibility, the common colour scheme is shading from red to green, where red indicates a higher (and therefore worse) implausibility value. This aligns with the common scheme used in previous papers (for example \cite{bower2010galaxy}, Figure 6), but can cause problems for colourblind users, particularly those with deuteranopia or protanopia. In any plot where this issue could arise, the \code{cb} option exists, which when set to \code{TRUE} plots instead on a two-colour gradient colour scheme that aligns with the tenets behind the \pkg{viridis} package \citep{garnier2021viridis}. Any other plots in the \pkg{hmer} package use either a single-colour scheme or one of the \pkg{viridis} colour schemes.
    }%
}

There are many other visualisation functions available within the \pkg{hmer} package: for instance, functionality to visually examine which variables are active for a collection of emulated outputs (\code{plot_actives()}); plotting tools to identify the most influential inputs to each output (\code{effect_strength()}); simple plotting functions to demonstrate output-input dependence based on either the emulator predictions or on the simulator output directly (\code{output_plot}, \code{behaviour_plot()}); and a pairs plot to simultaneously examine diagnostics and implausibility relative to their position in parameter space (\code{validation_pairs()})\footnote{For details of any of these, see the documentation (e.g., \code{?validation\_pairs}).}. We detail only two further emulator visualisation tools here, which deal with considering implausibility: the first is the \code{plot_lattice()} function and the second is the \code{space_removed()} function.

Despite its utility, a two-dimensional slice of parameter space does not give us a definitive picture of where acceptable matches might be found, and is highly dependent on the slice chosen. For instance, suppose the acceptable part of parameter space here lies in a small neighbourhood of $\alpha_{SI} = 0.45,\,\alpha_{IR} = 0.25,\,\alpha_{SR} = 0.01$ (i.e., the centre of the space in $\alpha_{SI}$ and $\alpha_{IR}$ but toward the lower end of the range for $\alpha_{SR}$), and we consider implausibility plotted over $\alpha_{SI}$ and $\alpha_{IR}$ with $\alpha_{SR}=0.045$. The plot thus produced would be unlikely to show the probable location of the non-implausible region, and in fact may suggest that a non-implausible region could exist away from the centre of the space since the variance may be higher at the edges of the space. We could produce multiple slices through the space quite easily in this example, but for higher-dimensional spaces this becomes infeasible. Instead, we may turn to the \code{plot_lattice()} command.

Taking as a minimum the emulators and the corresponding targets, the \code{plot_lattice()} function returns a grid of plots. Each plot represents a summary of the implausibility over the entire space, projected in the relevant fashion:

\begin{itemize}
\item Minimum maximised implausibility (upper triangle): the maximum (or $n$th-maximum) implausibility is calculated across the space. We collect implausibilities based on the projection of the input points into the particular two-dimensional subspace of interest, and calculate the minimum of the ($n$th) maximum implausibilities. This quantity is plotted;
\item Two-dimensional optical depth (lower triangle): Points are collected in the same fashion as above, but we instead consider the proportion of points projected that have acceptable implausibility. A value of $1$ suggests that all projected points are acceptable; a value of $0$ suggests that none are;
\item One-dimensional optical depth (diagonal): Similar to the two-dimensional case but projecting to a single parameter direction. Each of these plots has y-axis range of $[0, 1]$.
\end{itemize}

The term `optical depth' corresponds to a particular conceptualisation of the non-implausible region. Let us assume that we have determined an implausibility cutoff that delineates acceptable and unacceptable points, and choose two parameter directions of interest. The optical depth is then defined as the thickness of the non-implausible space along the line of sight defined by picking fixed values for the two chosen parameters. This is easiest to visualise in three dimensions, as the optical depth for any pair of the chosen parameters is the density along a line in the remaining direction, but the concept applies in higher dimensions. This, combined with the minimum maximised implausibilities, give the outline of the non-implausible region as well as a measure of its density: for more details, see \cite{vernon2018bayesian}. Note that for a `true' description of optical depth, one would need to integrate over the $(d-2)$ directions not included in the plot for each pair of parameters; the efficiency of emulator evaluation means that in many applications we may approximate this integration by considering sufficiently many points in the $(d-2)$~dimensional region. For lower dimensional input spaces, we evaluate the emulators over a fine regular grid; for higher dimensional spaces where this becomes computationally infeasible even using the emulators, we generate a large Latin hypercube from which we infer the density for each pair of parameter values of interest.

\begin{figure}[h]
\centering
\includegraphics[width = 0.7\textwidth]{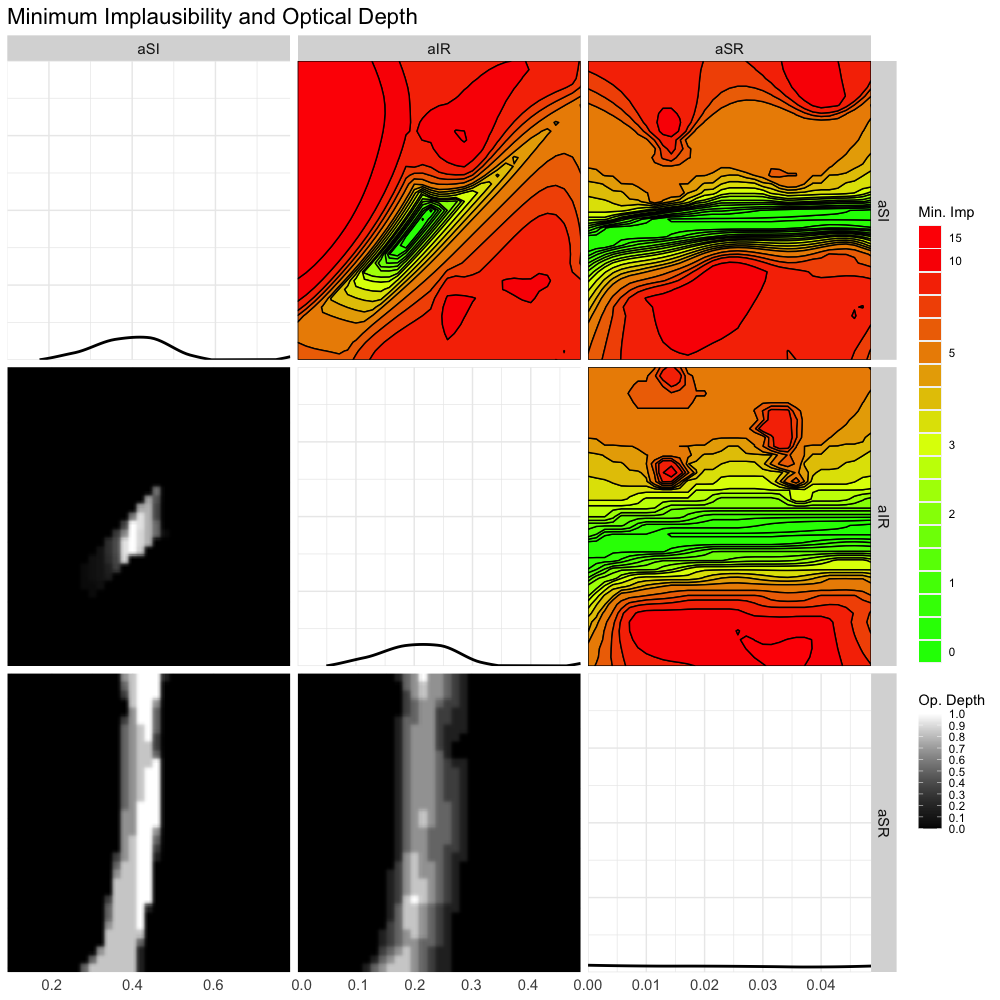}
\caption[Plot Lattice Output]{The result of a call to \code{plot\_lattice()}. Each plot is a particular combination of parameters; the upper diagonal provides minimum implausibility across the space; the lower diagonal the two-dimensional optical depth; the diagonal one-dimensional optical depth. We note that the combined effect of $\alpha_{SI}$ and $\alpha_{IR}$ is driving the identification of the non-implausible region.}
\label{fig:plotlattice}
\end{figure}

The use of \code{plot_lattice()} is as below, with the corresponding output in Figure~\ref{fig:plotlattice}.

\begin{CodeInput}
R> plot_lattice(ems_wave1, targets, ppd = 35)
\end{CodeInput}

This visualisation can be extremely helpful in visualising the overall acceptability of the space with respect to pairs of arguments, as well as identifying correlation structure between input parameters (for example, that between $\alpha_{SI}$ and $\alpha_{IR}$). The optical depth also gives us an indication of where in the projected space the strongest chance of finding an acceptable match lies: we can see a stronger signal toward the centre of the $\alpha_{SR}$ parameter range. In this particular example, the inclusion of the $\alpha_{SR}$ parameter in the plot provides little additional information at this wave, due to its subdominant impact on the emulated outputs (as remarked on in Section~\ref{sec:initmodel}); we can see this by the predominantly straight lines in those plots that include it. In more realistic higher dimensional problems such a combined plot of all parameter pairs can be crucial in understanding the structure of the non-implausible space -- for example, in a more complex model we may treat this set of waves of history matching as exploratory, choosing in a final assay to exclude $\alpha_{SR}$ from our varying parameters. In so doing, we may account for this parameter's exclusion with suitable internal discrepancy via a sensitivity analysis based on this ensemble of runs, providing this to the emulators via the \code{discrepancies} argument mentioned in Appendix~\ref{app:emcorr}. The speed of emulator evaluation makes such an exploratory analysis feasible, as we have been able to explore the full parameter space and make this inference with only a handful of simulator runs to guide us.

Finally, the choice of implausibility cut-off is not always straightforward to determine. For a single emulator's implausibility, we are interested in finding a suitable value $c$ such that
$$P\left(\vert\E_D[f(x)]-z\vert \ge c\sqrt{\VAR_D[f(x)] + \VAR[e] + \VAR[\epsilon]}\right) \le \alpha$$
holds for a chosen $\alpha$. We often appeal to Pukelsheim's $3\sigma$ rule, choosing $\alpha = 0.05$ and use $I=3$ as the cut-off for acceptability \citep{pukelsheim1994three}; however, for $n$th maximised implausibility over multiple outputs may not be so clear. If we can assume that $m$ of the $N$ outputs are independent, and under the assumption of normality, then a heuristic argument for a reasonable cut-off can be found in \cite{goldstein2013assessing}. However, these assumptions may not be valid, or may be hard to justify. Instead we can explore the robustness of various implausibility cut-off choices by examining the proportion of the current non-implausible region that will be removed by particular choices. The \code{space_removed()} function provides this insight, as well as allowing us to consider how changes to model discrepancy and observation error can affect the space removed.

At the most basic level, all that is required for the \code{space_removed()} function is the set of emulators and their corresponding targets: the below command computes the space removal with a grid of $20^3 = 8000$ points from the current non-implausible region, considering modifying the observational error to varying proportions of its current value. We can, therefore, choose more aggressive or more cautious cut-off values with appropriate insight into the effect these will have on reduction of the parameter space.

\begin{CodeInput}
R> space_removed(ems_wave1, targets, ppd = 20) +
     geom_vline(xintercept = 3, lty = 2)
\end{CodeInput}

\begin{figure}[h]
\centering
\includegraphics[width = 0.7\textwidth]{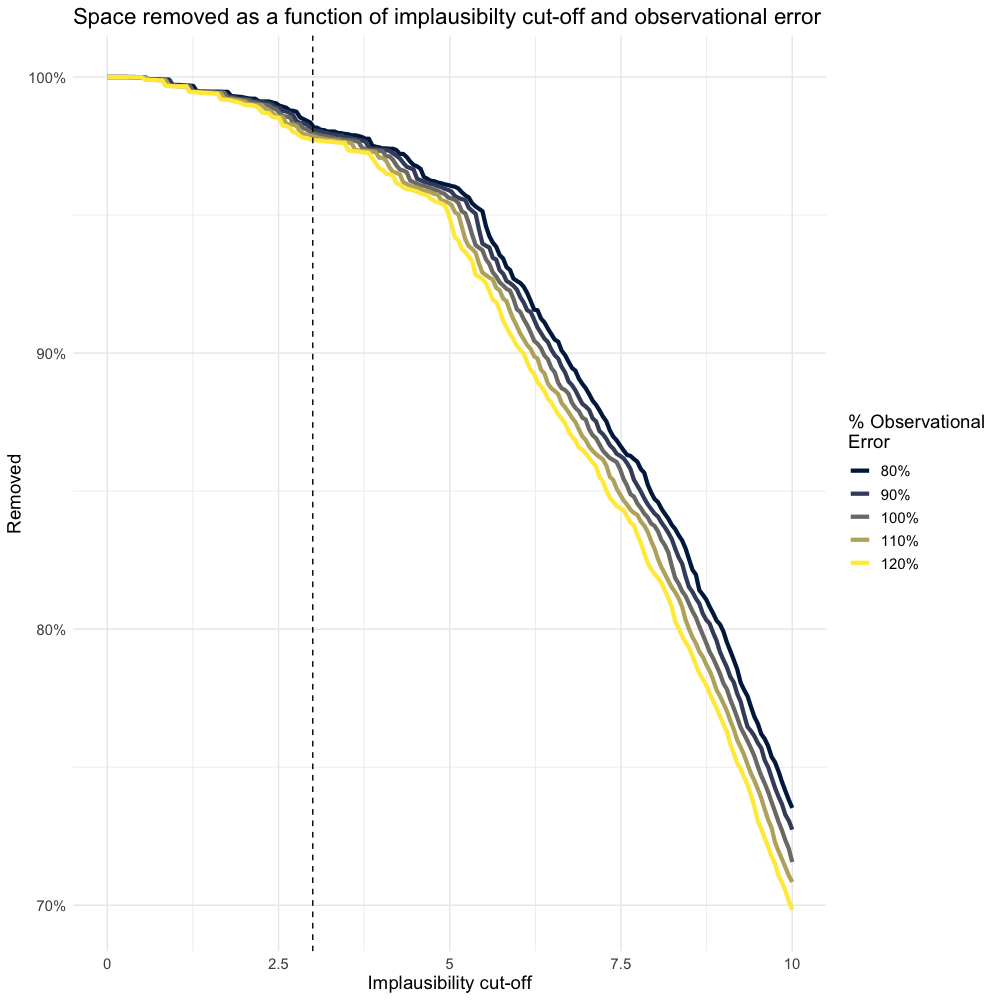}
\caption[Space removed]{The result of \code{space\_removed()} applied to our wave $1$ emulators. We can see that a cut-off of $I=3$ would be sufficient to rule out over $95\%$ of the current space, a figure that is relatively stable to changes in the choice of observational error. At higher cut-off values, the choice of observational error has a more pronounced effect.}
\label{fig:spaceremoved}
\end{figure}

The \code{space_removed()} function can also provide stratification of the space removed with respect to inflation and deflation of the prior emulator variances $\sigma_i^2$, the hyperparameters $\theta_i$, or the model discrepancies $\sigma_{m_i}$. This is controlled by the argument \code{modified}; the details of the stratification are determined by the \code{u_mod} argument which by default is \code{c(0.8, 0.9, 1, 1.1, 1.2)}. As with many of the plots in \pkg{hmer}, they can be augmented with additional \pkg{ggplot2} objects; in Figure~\ref{fig:spaceremoved} we have added a vertical line at $I=3$ using the \code{geom_vline()} function.

\begin{CodeInput}
R> # An example using the optional parameters
R> space_removed(ems_wave1, targets, modified = 'var', ppd = 20,
    u_mod = seq(0.5, 1.5, by = 0.1))
\end{CodeInput}

The collection of visualisation tools available within the \pkg{hmer} package allow us to gain a deeper understanding of the structure of the simulator, the input space, and the expected size and shape of the non-implausible region we expect to obtain from point proposals. For complex models where evaluations are expensive, this can be an invaluable tool for supplementing and augmenting our knowledge about the model. In high-dimensional models (in input and/or output dimension), some of the visualisations require a moderate amount of computational time; however, this time should be weighed against the equivalent time it would take to perform simulator runs at thousands of parameter sets and the ensuing analysis provided automatically by these functions. In all but the fastest of computer models, the emulators allow for an exploration of the model behaviour across the parameter space that simply would not be possible otherwise: an emulator evaluation at a parameter set takes between $10^{-6}$  to $10^{-2}$ seconds, whereas simulators on which emulation has been applied have varied in evaluation time between seconds \citep{scarponi2023tbhiv}, hours \citep{andrianakis2015bayesian}, and days \citep{vernon2014galaxy}, representing an efficiency gain of $10^3$ to $10^{10}$ compared with running the simulator.

\subsection{Proposing points and inspecting results}

The point proposal stage, for creating a design of points for the next wave, consists of a single call to the function \code{generate_new_design()}. As with \code{emulator_from_data()}, there are a multitude of optional arguments which are described in the associated help file; we detail the most salient of those arguments here.

Here, we apply the default usage discussed in Section~\ref{sec:propose}.

\begin{CodeChunk}
\begin{CodeInput}
R> proposal1 <- generate_new_design(ems_wave1, 500, targets)
\end{CodeInput}
\begin{CodeOutput}
[1] "Proposing from LHS..."
[1] "102 initial valid points generated for I=3"
[1] "Performing line sampling..."
[1] "Line sampling generated 40 more points."
[1] "Performing importance sampling..."
[1] "Importance sampling generated 474 more points."
[1] "Resample 1"
[1] "Performing line sampling..."
[1] "Line sampling generated 40 more points."
[1] "Performing importance sampling..."
[1] "Importance sampling generated 385 more points."
[1] "Selecting final points using maximin criterion..."
\end{CodeOutput}
\end{CodeChunk}

The function provides status updates as it proceeds through the stages of point proposal, and returns a \code{data.frame} of the points thus chosen\footnote{In this case we requested $500$ points, which would certainly be far more points than we would need for a second wave of history matching for this particular simulator. However, there is little computational disadvantage to doing so, given the speed of emulator evaluation.}. These points can be passed to the simulator as a representative sample of the non-implausible space $\mathcal{X}_1$, and we may start the next wave of history matching. We may also plot these in whatever fashion we desire: the native \proglang{R} \code{plot()} command produces an adequate pairs plot of inputs, or we can use \code{plot_wrap()} to plot the proposed parameter sets with respect to other ranges: for example, plotting the points of $\mathcal{X}_k$ with respect to the minimum enclosing hyperrectangle of $\mathcal{X}_{k-1}$ to inspect the shrinkage of parameter space.

\begin{CodeInput}
R> plot_wrap(proposal1, ranges)
\end{CodeInput}

\begin{figure}[h]
\centering
\includegraphics[width = 0.7\textwidth]{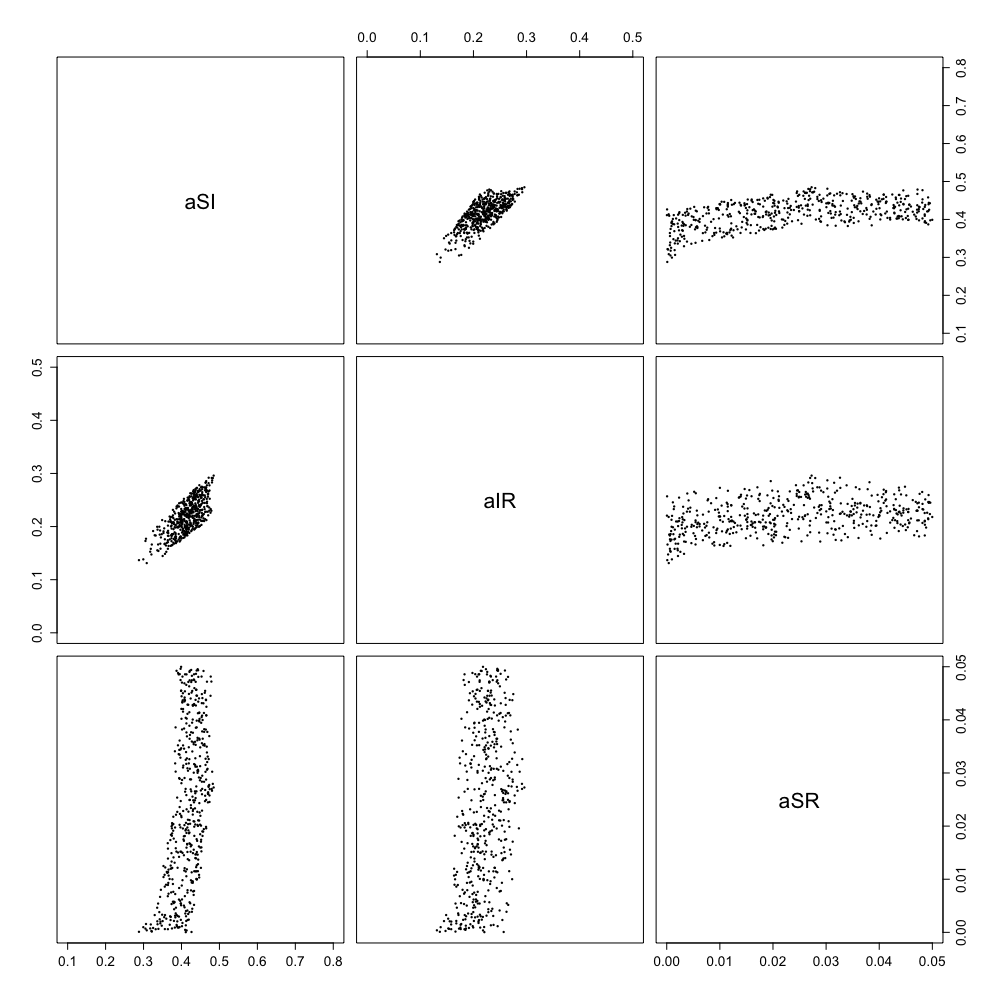}
\caption[Plotting Proposal]{A simple plot of the points proposed at wave $1$, with respect to the original ranges of the parameters. We can see the parameter restrictions, as well as the correlation between $\alpha_{SI}$ and $\alpha_{IR}$.}
\label{fig:plotwrap}
\end{figure}

\subsection{Visualising the simulator space}

After a few waves of history matching, we will have gained an evolving picture of the structure of the non-implausible space as well as the behaviour of the corresponding simulator runs. It is often instructive to look at this changing picture, so as to gain an understanding into which parameters are driving the reduction of non-implausible space, how the outputs react to changing parameter values, and indeed the progress towards matching all targets. To this end, a set of multi-wave visualisation tools are available: \code{wave_values()}, \code{wave_points()}, \code{wave_dependencies()}, and \code{simulator_plot()}. The \code{wave} prefixed functions return a grid of plots in output, input, or input-output space (respectively), allowing us to look at the behaviour and evolution of the space over multiple waves. The final plot is a simple heuristic plot that allows us to evaluate the suitability of simulator runs over the ensemble of points at each wave. Here, we demonstrate the use of \code{wave_values()} and \code{simulator_plot()}; the equivalent output of \code{wave_points()} is used throughout the paper and can be found in the Supplementary Material.

These functions are most effective when we have access to a few waves of history matching. One could quite easily replicate the steps demonstrated above (using \code{get_res} to obtain simulator runs at the end of each wave), but here we will use an \pkg{hmer} dataset which contains sets of points sampled from $\mathcal{X}_i$ for $i=0,\dots,3$, namely \code{SIRMultiWaveData}. This dataset consists of a four-element list, where each element is the relevant \code{data.frame} of points. The \code{wave_values()} function, as mentioned, returns a grid of plots (in the fashion of a pairs plot) with axes corresponding to the simulator outputs. The lower diagonal shows the values of the simulator runs, coloured by which wave their generating point was proposed at, as well as an overlay of the target bounds for the two targets in question\footnote{For targets that have been defined in terms of a value and a sigma, the bounds correspond to $3\sigma$ below and above the value.}. The upper diagonal shows the same plot but `zoomed in' to around the target bounds, in order to see the structure in the region of interest. The diagonal shows density plots for each individual output; here the target bounds are simply vertical lines. A simple code chunk is shown below, along with the resulting plot in Figure~\ref{fig:wavevalues}. Here, we have used the default arguments to the function, aside from using \code{l_wid} to slightly thin the target bound lines; options for excluding particular waves, controlling the level of `zoom', and plotting particular outputs, among others, are available.

\begin{CodeInput}
R> wave_values(SIRMultiWaveData, targets, l_wid = 0.8)
\end{CodeInput}

\begin{figure}[h]
\centering
\includegraphics[width = 0.7\textwidth]{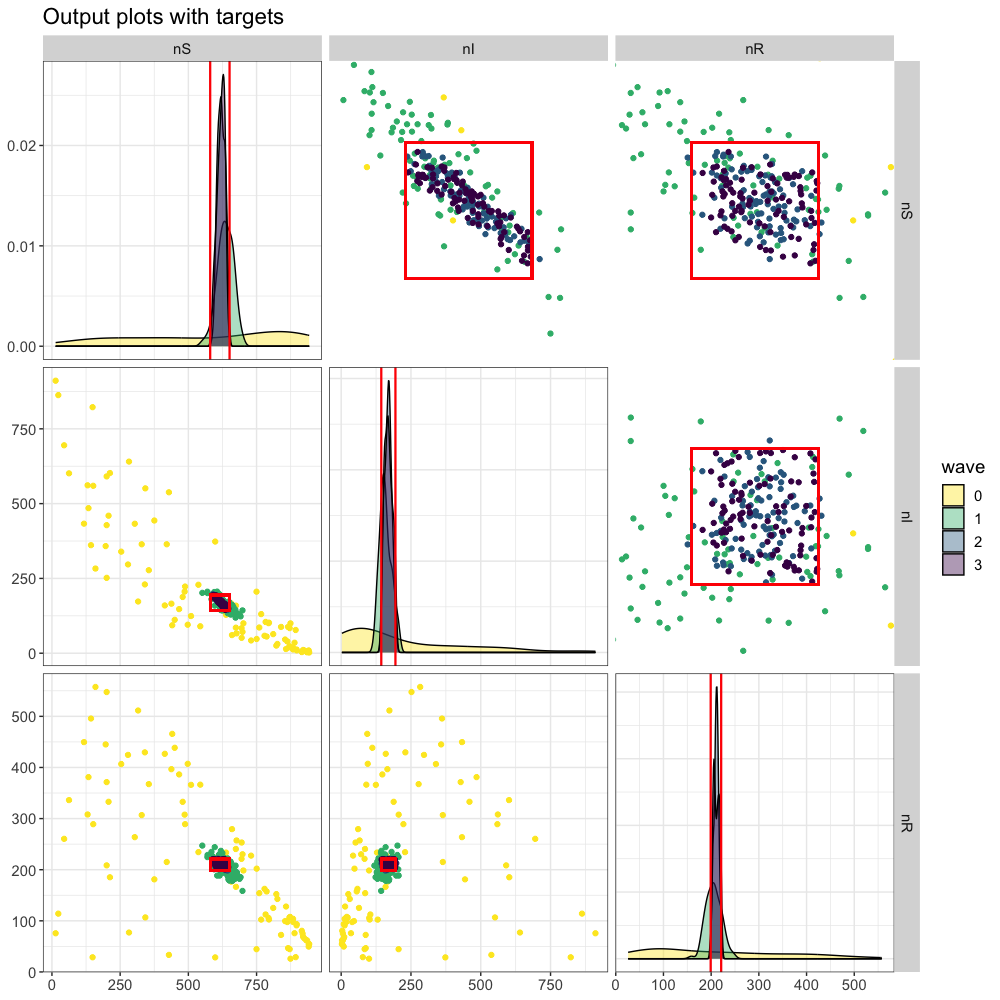}
\caption[Wave values]{The result of \code{wave\_values()} applied to the \code{SIRMultiWaveData} dataset. As we proceed through the waves, we see that more and more of the proposed points are lying within the bounds of the targets, as well as a strengthening correlation between $nS$ and $nI$.}
\label{fig:wavevalues}
\end{figure}

As well as providing an insight into the dependencies of the simulator, we may also consider this a barometer of how many waves we require to generate a large sample from our final non-implausible region. Here, we can see that most of the points proposed at wave $3$ lie within the bounds of every target, and so we could determine that further waves are unnecessary. This is backed up by a quick inspection of the emulator uncertainties at the final wave (accessible via \code{load("SIRMultiWaveEmulators")}; a list of three sets of trained emulators).

\begin{CodeChunk}
\begin{CodeInput}
R> sapply(SIRMultiWaveEmulators[[3]], function(x) x$u_sigma)
\end{CodeInput}
\begin{CodeOutput}
      nS       nI       nR 
1.4776221 2.1416256 0.5041151 
\end{CodeOutput}
\end{CodeChunk}

Comparison of these with the uncertainties on the observations themselves suggests that the emulator uncertainty is vastly subdominant to that of the observation, and subsequent waves of emulation will not substantially improve on the emulators from wave $3$. This is partly because of the simplicity of the toy model we are using (both in terms of the model dynamics itself and the low parameter and output dimension) and this three-wave completion should not be viewed as a `standard'; whereas some models can be matched to data within $3$ to $5$ waves \citep{bower2010galaxy, craig1997pressure}, the necessity of performing around $10$ waves \citep{andrianakis2015bayesian} or even closer to $20$ \citep{scarponi2023tbhiv} is not uncommon. Nevertheless, the techniques to evaluate the progress of the history matching remain true; plots such as \code{wave_values()} give us an easily interpretable measure for the status of the history match.	

One thing that cannot be (directly) determined from the \code{wave_values()} plots is whether a particular parameter set matches all targets. The \code{simulator_plot()} function provides a visualisation of the most fundamental question in the history matching process: are there any points that match to all outputs?

\begin{CodeInput}
R> simulator_plot(SIRMultiWaveData, targets, barcol = 'black')
\end{CodeInput}

\begin{figure}[!h]
\centering
\includegraphics[width = 0.5\textwidth]{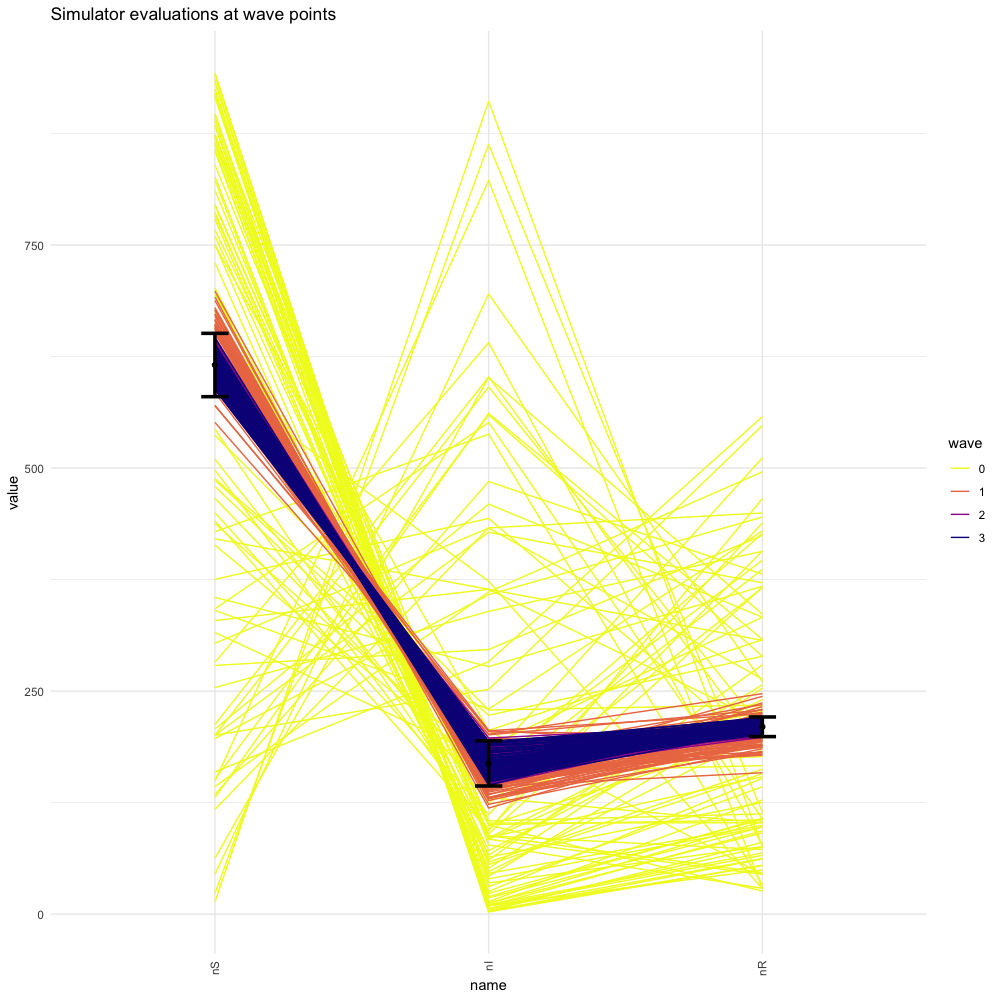}
\caption[Simulator Plot]{All `trajectories' across the outputs, for all waves, plotted using \code{simulator\_plot()}. The wild fitting behaviour in our wave $0$ (yellow) quickly settles down as the emulators propose points, until at wave $3$ (blue) practically all trajectories pass through the targets.}
\label{fig:simplot}
\end{figure}

In this case, the plot in Figure~\ref{fig:simplot} indicates to us the quality of the points proposed at wave 3. In more complex systems where a match is harder (or impossible) to obtain, this plot in conjunction with the \code{wave_...} plots can give an insight into which outputs are being missed, whether there are conflicts between outputs, or how close we are to obtaining a fit to all targets. Such an analysis, using these visualisation tools, was performed in \cite{scarponi2023tbhiv}, where the inability to find matches to observational data in a small number of cases could be traced to conflicts between particular outputs or potential misspecification in the observations.

In this section we have detailed the main stages of emulation and history matching using the \pkg{hmer} package. This introduction is by no means comprehensive, but aims to highlight the most important functions and visualisation tools. More involved examples of using the \pkg{hmer} package can be found in the vignettes, including lower- and higher-dimensional examples, demonstrations of more visualisation tools, and a brief exploration of some of the advanced techniques we are about to touch upon.

\section{Variance emulation and prototypical emulators}\label{sec:advanced}

\subsection{Variance emulation}\label{sec:vem}
The flexibility of the Bayes linear emulation approach allows us to extend the framework beyond that of simple emulation of deterministic systems. In this section, we primarily discuss an aspect of more advanced functionality in the \pkg{hmer} package --- namely, the emulation of stochastic systems using hierarchical emulation.

The simple example described in Section~\ref{sec:ex} was defined by a set of differential equations, which naturally (up to machine precision in the ODE solver of choice) result in the same output for repeated runs of the same input parameters. Complex models do not always possess this \emph{deterministic} behaviour: in particular, in epidemiology, a stochastic compartmental model or an agent-based model (ABM) can be used, where transition between states in the system is probabilistic \citep{wilkinson2018stochastic}. As a result, repeated simulator runs at the same parameter values (which we term \emph{realisations}) will result in different values of the outputs. Such a simulator is referred to as \emph{stochastic}.

Stochastic simulators can (and often do) exhibit heteroskedasticity, where the variance across the parameter space can be wildly different due to the variability in the model. For example, consider a simple Birth-Death model where individuals are born at a rate $\lambda$, and die at a rate $\mu$. For low values of $(\lambda, \mu)$, the system changes very little over time and the variability is minimal (the limit where $\lambda, \mu\to 0$ of course gives no variability whatsoever); for high values of $(\lambda, \mu)$ the variability over multiple repetitions of the simulator is much higher.

For such systems, we may apply a hierarchical approach to emulation, leveraging the flexibility of the Bayes linear emulation approach \citep{cumming2010bayes}. Suppose we have a set of simulator outputs at points $(x^{(1)}, x^{(2)}, \dots, x^{(n)})$ where, for each point $x^{(j)}$, we have repeated the simulator evaluation $N_j$ times. We therefore have a collection of $\sum_j N_j$ simulator runs, which can be categorised by where in parameter space they were evaluated. The $N_j$ need not be the same at each point: for instance, where a simulator is slower to run in a part of parameter space we may perform fewer runs than at a more `well-behaved' region of parameter space. This possible distinction between information at points is handled by the package.

From the collections of outputs we can determine a collection of sample means and sample variances, indexed by $x^{(j)}$. The key stage in emulating this system is to treat the sample variances, which we denote $s^2_i(x^{(j)})$, obtained as an output in their own right and create an emulator for the variance of the stochastic simulator itself. This requires an understanding of, or sensible priors for, the fourth-order quantities of the system; since our `expectation' $\E[s^2_i(x)]$ is the variance of the system, the `variance' $\VAR[s^2_i(x)]$ is the variance of the variance. There are various ways to decide on sensible priors for the relevant quantities \cite[Ch.~8]{goldstein2007bayes} which provide a set of emulators that can offer predictions as to the stochasticity of the simulator across the non-implausible region\footnote{If no expert prior knowledge is available for these quantities, \pkg{hmer} will determine these as part of the emulator training process.}. As with all such emulators, they know the level of stochasticity at training points, and provide a statistical approximation of it at unseen points.

Of course, we want to explore the space of simulator outputs, not of its variability. For example, the primary interest is often in the examination of the mean surface of the stochastic model. However, the variance emulators trained above can help to correct for any uncertainty induced by the fact that, due to finite sample size, we cannot observe the `true' underlying mean process from a collection of realisations of the stochastic simulator. The use of variance emulation therefore allows us to train much more representative emulators to the mean surface which, moreover, have an intrinsic understanding of the underlying stochastic process.

Training such emulators is a multi-step process, naturally, but aside from some considerations has much the same structure as the training process for deterministic emulators. To that end, the function \code{emulator_from_data} also provides the means by which this hierarchical structure can be determined, constructed, and adjusted with respect to data. It takes the same collection of arguments, with two caveats: firstly, the data provided to \code{input_data} should be \emph{unaggregated}, comprising data from all realisations from all input parameter combinations; secondly, we provide the argument \code{emulator_type = "variance"}.

The estimates of the fourth-order quantities are determined, dependent on the nature of the data available, and the correction for sample quantities is incorporated. Unlike the output of \code{emulator_from_data}, this function returns a nested list of emulators: one set named \code{variance} and one named \code{expectation}\footnote{A note on terms here: we use `expectation' to denote the mean emulators in order to avoid possible conflicts with \proglang{R}'s native \code{mean} function.}, corresponding to the variance and mean emulators for the outputs, respectively. Once trained, this collection of emulators can be passed as an argument to any of the usual functions as appropriate: in particular, the core functions handle the nested collection of emulators in the manner appropriate to their usage.

We briefly show the results of training emulators to stochastic data here, using the stochastic equivalent of the SIRS model described in Section~\ref{sec:ex}, where results are generated using the Gillespie algorithm. Much of the \pkg{hmer} code is unchanged and does not bear repetition (one may find sample code within the Supplementary Material, snippets of which we detail below); we merely highlight the comparative ease of use of the \pkg{hmer} package when dealing with stochastic simulators, and highlight a few key differences between stochastic and deterministic emulation.

\begin{CodeInput}
R> stoch_emulators <- emulator_from_data(training_stoch, names(targets),
    ranges, emulator_type = "variance")
R> emulator_plot(stoch_emulators$variance[c('nS', 'nR')])
\end{CodeInput}

\begin{figure}[!h]
\centering
\includegraphics[width = 0.85\textwidth]{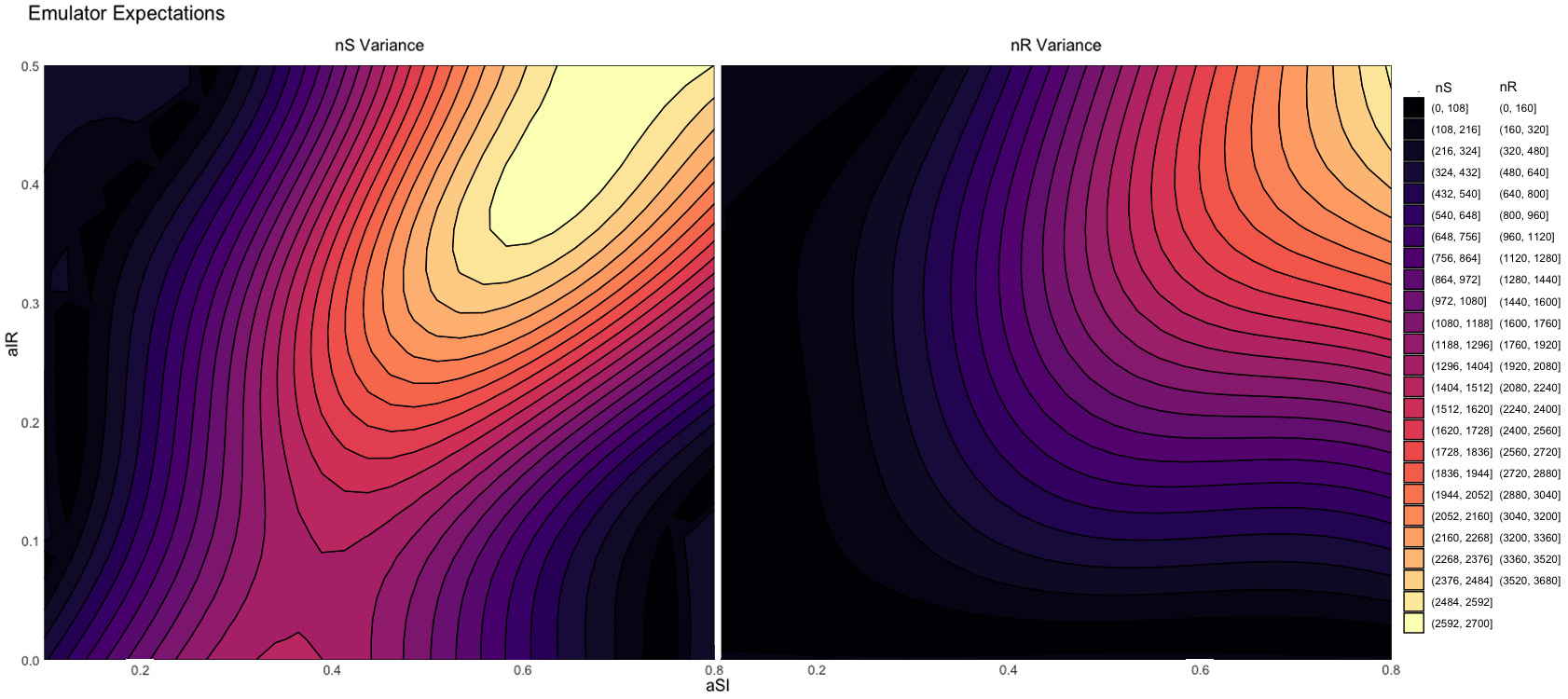}
\caption[Stochastic Plot]{Emulator plots of expectation of variance emulators.}
\label{fig:stochplots}
\end{figure}

Figure~\ref{fig:stochplots} shows that the stochasticity is very different across the space: in particular the model is far more variable for high values of $\alpha_{SI}$, as one would expect. The increased stochasticity detected by the variance emulators informs the mean emulators, while not dominating the determination due to training points. The effect of hierarchical emulation is most clear when we evaluate the emulator uncertainties at training points:

\begin{CodeChunk}
\begin{CodeInput}
R> stoch_emulators$expectation$nI$get_cov(train_sample)
\end{CodeInput}
\begin{CodeOutput}
[1] 3.8808528  4.4911750 3.9082331  0.5227077  3.7521496
[6] 3.3824107 4.1191526 2.9284490 4.0294971 0.5732920
\end{CodeOutput}
\end{CodeChunk}

While for deterministic systems the uncertainty at points used to train the emulators identically vanishes, this is not the case for a stochastic emulator. This is a consequence of the fact that the emulators are designed so as to predict the `true' mean response of the simulator that would be seen if we could perform infinitely many realisations; since we only have finitely many realisations at each training point the emulators automatically incorporate this uncertainty due to finite sample size.

Validation of emulators proceeds in the same way with \code{validation\_diagnostics()}, with the understanding that the validation set again comprises the unaggregated results from the simulator; the process of generating new runs with \code{generate\_new\_design()} is also unchanged from a code perspective.

\begin{CodeInput}
R> validation_stoch <- validation_diagnostics(stoch_emulators, targets, 
    stoch_valid)
R> stoch_points <- generate_new_design(stoch_emulators, 90, targets)
\end{CodeInput}

\begin{figure}[h]
\centering
\begin{subfigure}{0.45\textwidth}
\includegraphics[width = \textwidth]{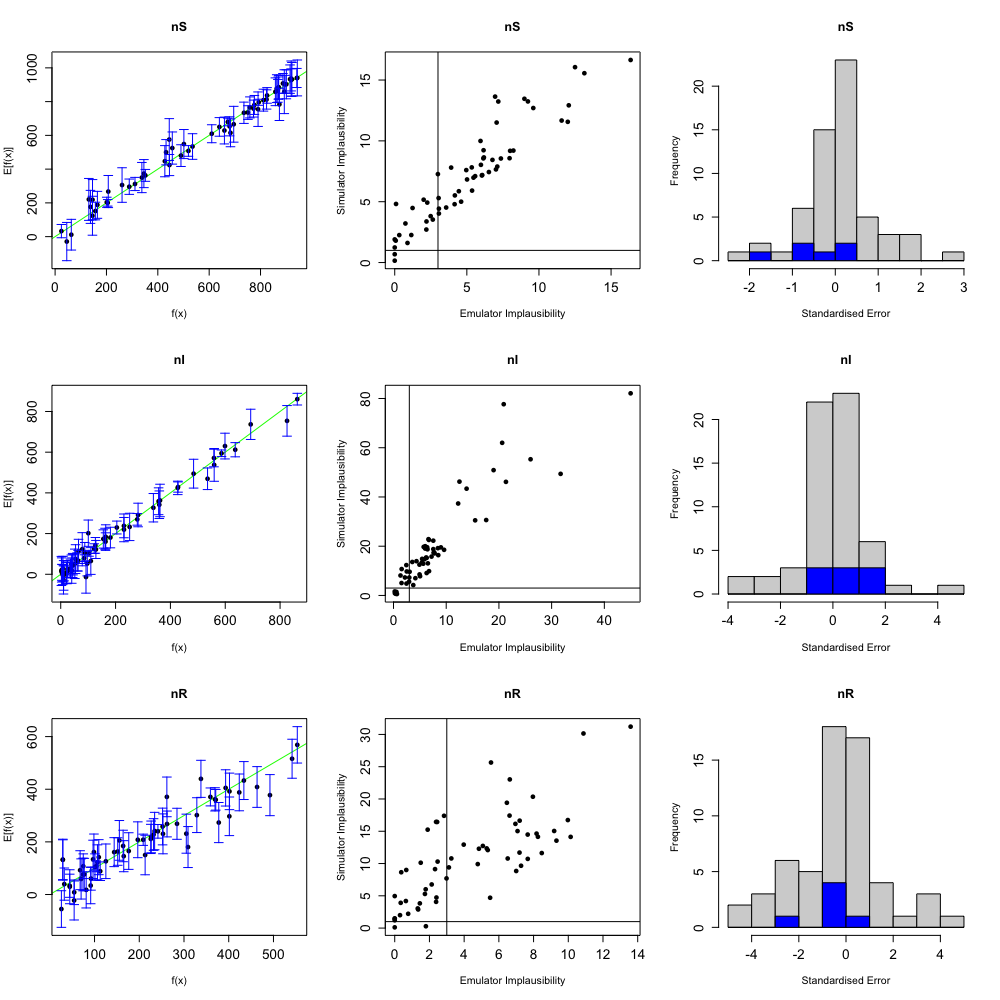}
\end{subfigure}
\begin{subfigure}{0.49\textwidth}
\includegraphics[width = \textwidth]{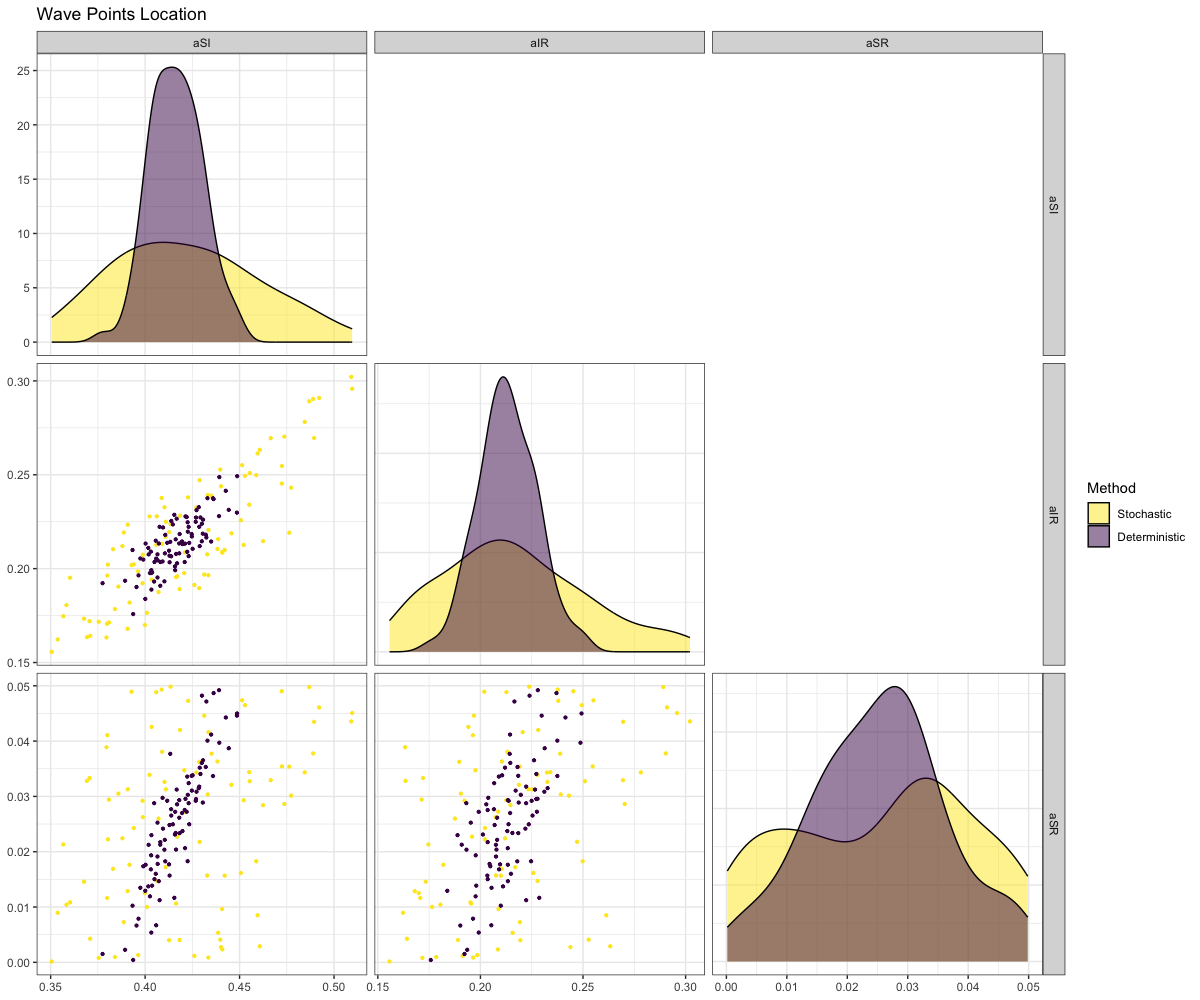}
\end{subfigure}
\caption[Stochastic Emulation Results]{Diagnostics for variance emulators (left), and a comparison of point proposals for the stochastic (yellow) and deterministic (purple) simulators.}
\label{fig:stochvdeterm}
\end{figure}

Figure~\ref{fig:stochvdeterm} shows the result of proposing points after two waves of emulation and history matching of both the deterministic model introduced in \ref{sec:ex} and its stochastic equivalent introduced here. We can see that the stochastic NROY space is a superset of its deterministic counterpart: this is as we would expect since the calculation of non-implausibility takes into account the stochasticity of the model. In particular, we know from Figure~\ref{fig:stochplots} that the stochasticity is strongest for high values of $\alpha_{SI}$ and $\alpha_{IR}$: this is borne out by the fact that this is the region of non-implausible space in Figure~\ref{fig:stochvdeterm} that is furthest from the deterministic equivalent. The difference between the two proposals, even for a simple model such as the SIRS model under consideration, highlights the importance of robustly incorporating the effect of stochasticity into a calibration process: without it, we run the risk of ignoring parts of parameter space that could actually represent the real-world behaviour we seek to match to.

This machinery allows us to accurately emulate heavily stochastic systems with minimal modification to the history matching workflow described in Section~\ref{sec:hm}. If our only interest is in matching to the means of the stochastic outputs, then the emulators for the variance contribute only at the stage of building the mean emulators and do not complicate the procedure of history matching; if in fact we have observations of the real-life variability of the system in question then we can use the variance emulators alongside those for the mean to find parts of parameter space with the expected output behaviour and the expected variation, allowing for more powerful and efficient waves of history matching.

\subsection{``Prototyping'' emulation}\label{sec:proto}

Throughout the paper, we have presented emulation and history matching as an inseparable process where one requires the use of the other. However, the use of Bayes linear emulation is not solely useful for history matching; nor does history matching require Bayes linear emulators. For example, one may consider optimal design using Bayes linear emulators \citep{jones2016bayes}, thereby using a different metric for `suitability' than the standard approach in history matching, or use the framework of history matching and point proposal with Gaussian process emulators \citep{audouin2022modeling} Here we briefly discuss ways in which a user may separate the two procedures.

In the former situation, the \code{generate_new_design} function can be used with arbitrary metrics via the \code{accept_measure} argument in \code{opts}. The syntax for such a function is as follows.

\begin{CodeInput}
R> custom_func <- function(ems, x, z, cutoff, ...)
\end{CodeInput}

The output of this function should follow the format of implausibility, returning either a boolean or a numeric for a parameter set. This allows a generic measure of suitability when considering proposals, regardless of how the emulators are to be used. This also allows for a semi-multivariate approach to emulation, where we may define the implausibility using a Mahalanobis distance rather than using an nth maximised implausibility, or a composite measure where (say) we require maximum implausibility no greater than $4$ and second-maximum no greater than $2.5$.

In the latter situation, the \code{Proto_emulator} object allows for generic functions to be treated as if they were emulators, to leverage visualisation, point proposal and history matching mechanisms within \pkg{hmer}. This creates \pkg{R6} objects akin to \code{Emulator} objects; the minimal syntax of the constructor is

\begin{CodeInput}
R> Proto_emulator$new(ranges, output_name, predict_func, variance_func)
\end{CodeInput}

This structure allows the majority of functionality within the package to be used with other predictive methods - for instance, in emulating likelihoods where the structure of a Bayes linear emulator is not necessarily useful or appropriate.

The functionality for both of these use-cases can be found in more detail in the \pkg{hmer} documentation. The inclusion of these options in the package is an appreciation of the value of other approaches, depending on the application, while still wanting to make accessible the diagnostic and visualisation tools and other relevant aspects of the package functionality.

\section{Conclusion and discussion}\label{sec:conclude}

The \pkg{hmer} package provides an accessible means of applying the powerful techniques of Bayes linear emulation and history matching to complex, often computationally intensive, simulations of real-world processes. The framework has been employed to great success in a variety of modelling situations, and the \pkg{hmer} package itself has been used by epidemiologists to find matches to TB and HIV simulators of varying complexity and evaluation time \citep{scarponi2023tbhiv, clark2022tbhiv}. The core functionality of the package is designed to require, by default, as little understanding of the mathematical machinery that underpins emulation as is possible, while still providing the flexibility for advanced users to create bespoke emulators for their particular circumstance using expert judgements for prior specifications.

In this article, we have detailed the main, `front-facing', functions of the package, with which a user can follow the process of training and validating emulators and propose appropriate representative collections of points from the resulting non-implausible space. We have also detailed, by demonstration, a sample of the visualisations that can be derived from the computationally fast emulators to gain insight into the structure of the simulator they purport to replicate. The toy model we have used to demonstrate these features is of simplistic form, but the principles and processes apply equally to simulators whose run-time per evaluation is of the order of days, rather than seconds \citep{williamson2013history, vernon2022bayesianjune, vernon2014galaxy}. The comparatively small number of parameter sets required to train the emulator minimises the computational load for slow simulators and allows a thorough investigation of the non-implausible space at every stage, resulting in a potentially very large collection of suitable points without constant recourse to the simulator. Points generated from the non-implausible space can be used to fully furnish and investigate the space of acceptable matches to data in a number of different ways, which we have outlined.

The \pkg{hmer} package is model-agnostic and code-agnostic, requiring only that the outcome of multiple simulator runs can be collated into a \code{data.frame} (such an object can be easily obtained from, say, a \code{csv} file, often requiring minimal user-time to produce). We have placed particular emphasis on epidemiological models and disease modelling, due to the specific features of such models and the obvious advantages some aspects of the package can offer; however, all techniques presented in the package have potential applicability across all fields of computer modelling. Many of the techniques employed here have existed in the literature for some time but remained out of reach of a modeller without a deep understanding of the emulation framework. The \pkg{hmer} package is designed with the intention of removing that barrier to usage.

Nevertheless, emulation is not a panacea for all difficulties in modelling. To best utilise the inherent uncertainty structure, it is incumbent on the user to think carefully about their simulator, any discrepancies between it and reality, and the quality and accuracy of the observational data that they wish to match to. The framework provided for emulation is permissive with regards to the specification of these quantities; this does not mean that it should be ignored if one wishes to comprehensively determine the space of acceptable matches, subject to all the uncertainties that separate the simulator from the real-world process.

Similarly, the automated process of emulator training and even the default choices for implausibility cut-offs, composite measures, and point sampling methods are designed to best accommodate as large a class of models as is possible. The scope of models and systems that these processes could be applied to is sufficiently diverse so as to preclude any unified automated approach to emulation and history matching, and the user should see emulation as an interactive process that allows them to learn about their simulator, strengths and drawbacks, while exploring the parameter space of interest. Here, we briefly discuss a few situations where user-interaction can vastly improve the quality and efficiency of a history match.

\begin{itemize}
\item Expert knowledge about a simulator can make a big difference to the quality of emulators. The choice of initial parameter space $\mathcal{X}_0$ can hugely impact the ability of the emulators to accurately represent the simulator, as can the choice of points within that parameter space upon which to train the emulators. If we anticipate that interesting behaviour should occur in particular regions of $\mathcal{X}_0$, then we should ensure that the emulators are privy to that information via judicious choice of training points. This becomes especially pronounced in high-dimensional models, where to fully represent the behaviour across the parameter space with a reasonable number of simulator evaluations is near-impossible. In such cases, more advanced `border-block' designs can be used \citep{cumming2009small}.

\item The process of training emulators is fast and there is little disadvantage to creating a collection of emulators and, in light of emulator diagnostics, retraining with additional points in parameter space designed to overcome any shortcomings identified. Similarly, the choice of targets can hugely impact the inferential power of the emulators; choosing which outputs to emulate at a given wave is currently a determination for the user, as well as whether there exist any sensible transformations of the data that highlight the structure of the simulator.

\item The default behaviour of \code{generate_new_design} will, for most non-implausible regions, produce a well-spread representative sample of the space. However, there are circumstances under which the algorithm described in Section~\ref{sec:propose} will be ill-suited to the task. Two obvious examples are when the non-implausible space $\mathcal{X}_{k+1}$ is many orders of magnitude smaller than its superset, $\mathcal{X}_k$, or when $\mathcal{X}_{k+1}$ consists of differently sized disconnected regions. In both cases, the default \code{generate_new_design} process may fail to pick up on the small regions of $\mathcal{X}_{k+1}$ or indeed fail to find the space at all. In such circumstances, it may be helpful to consider moving to a less restrictive composite measure of implausibility, relaxing the cut-off, or identifying the most restrictive outputs and removing them from this wave. In extremis, the function \code{idemc} exists as a computationally intensive search of the space which operates akin to Evolutionary Monte Carlo methods \citep{williamson2013implausibility}. This should be viewed as an approach of last-resort, coming after an investigation of the space, the emulators, and our own intuition about the behaviour of the simulator and of the observational data we possess.
\end{itemize}
In future updates to the \pkg{hmer} package we plan to enhance the functionality to make surmounting some of the issues mentioned above easier, as well as introducing new methodology for finding acceptable fits and leveraging the additional information that can be gained from an inspection of the emulator structure. For example, for models that are `easy' to match to (where the regression surface accounts for a large part of the behaviour across the space), it can be helpful to use the properties of the regression surface to heuristically identify promising regions of parameter space as well as diagnose regions that are underrepresented in training; we may use the active variable structure to `cluster' emulators and their active variables and so operate on multiple reduced-dimension parameter spaces when proposing points \citep{cumming2010bayes}; for heavily stochastic systems where multiple states could be present (for instance, `bimodality' of outputs) it may be appropriate to emulate the states separately in order to obtain a good fit to our simulator; we may even emulate the derivative of the simulator at a point with or without further information or training runs in order to identify best directions of approach for finding good parameter sets. These examples of enhanced functionality are present in some form in the \pkg{hmer} package, and will be further improved as the package evolves.

The \pkg{hmer} package is designed to allow a robust and careful analysis of a simulator, as well as identify regions of parameter space that produce an acceptable fit to observed data. History matching with emulation is a powerful tool in performing this delicate task, and our aim has been to provide a means by which modellers can implement these methods with a minimum of statistical involvement, while allowing a deeper inspection into the structure of the emulation for those that wish to do so. During development we have seen considerable success in the use of \pkg{hmer} to calibrate (both with and without direct user input) complex simulators \citep{scarponi2023tbhiv, clark2022tbhiv}, and we expect that the power that modellers and the wider community can leverage from these techniques, presented in a user-friendly form, will increase as the user-base and the scope of the package expands. Furthermore, having access to a tool that allows careful examination of models without recourse to expensive simulator runs will allow modellers to incorporate these techniques into model design and debugging, uncertainty quantification and forecasting, and model comparison more generally.

\section*{Acknowledgements}
AI and DS would like to acknowledge the support and funding provided by the Wellcome Trust grant 218261/Z/19/Z, as well as the valuable feedback and package testing provided by Rebecca Clark, Christinah Mukandavire, Chathika Weerasuriya, Arminder Deol, Roel Bakker, and all in the TB Modelling group at London School of Hygiene and Tropical Medicine. AI would also like to acknowledge the helpful comments of Dario Domingo on the final manuscript. TJM is supported by an Expanding Excellence in England (E3) award from Research England. IV would like to acknowledge the support of the UK Research and Innovation grant EP/W011956/1. RGW is funded by the Welcome Trust (218261/Z/19/Z), NIH (1R01AI147321-01), EDTCP (RIA208D-2505B), UK MRC (CCF17-7779 via SET Bloomsbury)), ESRC (ES/P009011/1), BMGF (OPP1084276, OPP1135288 \& INV-001754), and the WHO (2020/985800-0).

\bibliography{references}

\appendix
\section[The Emulator and Correlator Objects]{\code{Emulator} and \code{Correlator} objects}\label{app:emcorr}

\subsection[The Correlator Object]{The \code{Correlator} object}\label{sec:Correlator}

The \code{Correlator} object is an \pkg{R6} object which performs the function of $u(x_{A_i})+w_i(x)$, except with unit variance. Its constructor takes three arguments:

\begin{CodeInput}
R> Correlator$new(corr, hp, nug)
\end{CodeInput}
where \code{corr} is a string corresponding to the name of the desired correlation function, \code{hp} is a named list of hyperparameters for the chosen correlation function, and \code{nug} is the size of the nugget term (equivalent to $\delta_i$ in the emulator specification \eqref{eq:corr}). The default \code{Correlator} object is initialised as if by \code{Correlator$new("exp_sq", list(theta = 0.1), nug = 0)}.

The type of the first argument allows us to specify correlation functions either from a list of those available: in the \pkg{hmer} package are exponential-squared \code{"exp_sq"}, Mat\'ern \code{"matern"}, Ornstein-Uhlenbeck \code{"orn_uhl"}, and rational quadratic \code{"rat_quad"}. The \code{Correlator} object searches for an available function whose name matches the string provided to the \code{corr} argument, allowing a user to define their own correlation function \code{user_func(...)} and provide it to the \code{Correlator} object. The requirements of a user-defined function are detailed in the help file \code{?Correlator}. This approach to the emulator correlation structure allows for consistent hyperparameter estimation across any conceivable correlation functions within \code{emulator_from_data()}.

A \code{Correlator} object possesses a \code{print()} method, getters and setters for the hyperparameters, and most importantly a function that returns the correlation between two collections of points, \code{get_corr()}.

\subsection[The Emulator Object]{The \code{Emulator} Object}

The \code{Emulator} object is the central object in the \pkg{hmer} package. An individual emulator is an \pkg{R6} object: it is infrequently used directly but the constructor function for an \code{Emulator} object is

\begin{Code}
R> Emulator$new(basis_f, beta, u, ranges, ...)
\end{Code}

The details of each of these arguments is as follows:

\begin{itemize}
\item \code{basis_f}: a list of basis functions equivalent to $\{g_i(x)\}$. At a minimum, this should include the constant function \code{function(x) 1}, and the functions should be designed to act on a vector of numerics: for instance, the basis functions ${g_i(x)}:\,\mathbb{R}^2\to\mathbb{R}^3$, $(x, y)\mapsto(1, x, y)$ can be entered as
\begin{CodeInput}
R> bf <- c(function(x) 1, function(x) x[1], function(x) x[2])
\end{CodeInput}

\item \code{beta}: This provides the second-order specifications for the regression coefficients $\beta$. This consists of a named list \code{list(mu = ..., sigma = ...)} corresponding to the expectation vector and variance matrix for $\beta$, respectively. In the above example, we may have
\begin{CodeInput}
R> b <- list(mu = c(0.1,-1, 1), sigma = diag(0, nrow = 3))
\end{CodeInput}
were we to expect that the regression surface has the form $0.1-x+y$ with no variance in the regression surface. Note that this does not result in an emulator with no variance; the choice of zero-variance in the regression coefficients allows us to separate global and local behaviour, with all residual variance accounted for in the weakly stationary process \code{u}.
\item \code{u}: This represents the correlation structure, whose argument is a list \code{list(corr = ..., sigma = ...)}, where \code{corr} is the \code{Correlator} object and \code{sigma} the standard deviation given by $\sqrt{\VAR[u_i(x)]}$; for the above example we could decide on a `default' correlation structure with variance 4, giving
\begin{CodeInput}
R> u <- list(corr = Correlator$new(), sigma = 2)
\end{CodeInput}
\item \code{ranges}: A list of ranges of the parameters. This is required for a variety of reasons, not least because the \code{Emulator} object internally scales all points passed to it so that they are in the range $[-1,1]$, as well as to determine the range for proposing points for later waves. When provided to the \code{Emulator}, they should be in a named list where each element is a pair of numerics, representing the lower and upper bounds of the parameter.
\begin{CodeInput}
R> range <- list(x = c(-1, 1), y = (2, 4))
\end{CodeInput}
\end{itemize}

There are a number of optional arguments than can be supplied to the \code{Emulator} object: a \code{data.frame} named \code{data} if we wish to manually provide training runs to the emulator; a string \code{out_name} which identifies the output we are emulating; a vector of booleans \code{a_vars} (of length equal to the number of parameters) which determines which variables are active; and a list of two numerics \code{discrepancies(internal = ..., external = ...)} which encode the model discrepancies. All of these quantities can be provided with expert prior knowledge of the output, though in many cases this is not possible; we instead rely on the \code{emulator_from_data()} function described in Section~\ref{sec:emfromdata} to supply these details for us.

As an \pkg{R6} object, a constructed \code{Emulator} inherits the normal \code{clone()} function as well as possessing a custom \code{print()} and \code{plot()} statement. It also has a number of functions which allow the Bayes linear adjustment to be performed, and posterior predictions to be obtained. The function \code{adjust(data, out_name)} performs the Bayes linear adjustment relative to the data $D$ provided in \code{data}; the functions \code{get_exp(points)} and \code{get_cov(points, ...)} perform the duty of calculating the posterior expectation and covariance matrix for a collection of parameter sets passed to them; \code{implausibility(points, target)} gives the implausibility at a collection of points relative to an observational target value. In the normal course of history matching it is unlikely that these functions need to be called directly; the more front-facing usage that leverages these functions is demonstrated in Section~\ref{sec:ex}.

\end{document}